\documentclass[10pt, conference, letterpaper]{IEEEtran}
\usepackage{cite}
\usepackage{multicol}
\usepackage{flushend}
\usepackage{fancyhdr}
\usepackage{graphicx,latexsym}
\usepackage{epstopdf}
\usepackage{amssymb}
 \usepackage{booktabs,xcolor}
\usepackage{algpseudocode,algorithm,url,epsfig,times, subfigure, setspace, amsmath}
\begin{document}
\title{Topology Optimization for Galvanic Coupled Wireless Intra-body Communication}
\author{Meenupriya Swaminathan, Ufuk Muncuk and~Kaushik R. Chowdhury,~\IEEEmembership{Member,~IEEE}\\
Electrical and Computer Engineering Department, Northeastern University, Boston, MA 02115, USA. \\E-mail:\{meenu, muncuk.u, krc\}@ece.neu.edu.}
\maketitle
%\thanks{Meenupriya Swaminathan, Ufuk Muncuk and Kaushik R. Chowdhury are with the Electrical and Computer Engineering Department, Northeastern University, Boston, MA 02115 USA. email:\{meenu,umuncuk,krc\}@ece.neu.edu.}}
\maketitle
\IEEEpeerreviewmaketitle
\pagestyle{plain}
\setcounter{page}{1}
\thispagestyle{fancy}
\pagestyle{fancy}
\pagenumbering{arabic}
\begin{abstract}
Implanted sensors and actuators in the human body promise in-situ health monitoring and rapid advancements in personalized medicine. We propose a new paradigm where such implants may communicate wirelessly through a technique called as galvanic coupling, which uses weak electrical signals and the conduction properties of body tissues. While galvanic coupling overcomes the problem of massive absorption of RF waves in the body, the unique intra-body channel raises several questions on the topology of the implants and the external (i.e., on skin) data collection nodes. This paper makes the first contributions towards (i) building an energy-efficient topology through optimal placement of data collection points/relays using measurement-driven tissue channel models, and (ii) balancing the energy consumption over the entire implant network so that the application needs are met. We achieve this via a two-phase iterative clustering algorithm for the implants and formulate an optimization problem that decides the position of external data-gathering points. Our theoretical results are validated via simulations and experimental studies on real tissues, with demonstrated increase in the network lifetime. 
\end{abstract}
%\noindent\begin{keywords}
%Intra-body communication, galvanic coupling, channel model, circuit model, implanted sensors/actuators, tissue safety.
%\end{keywords}
\section{Introduction} \label{sec:intro}
\begin{figure}[t]
 \centering
 \vspace{-2mm}
\includegraphics[width=8.5cm,height=4.2cm]{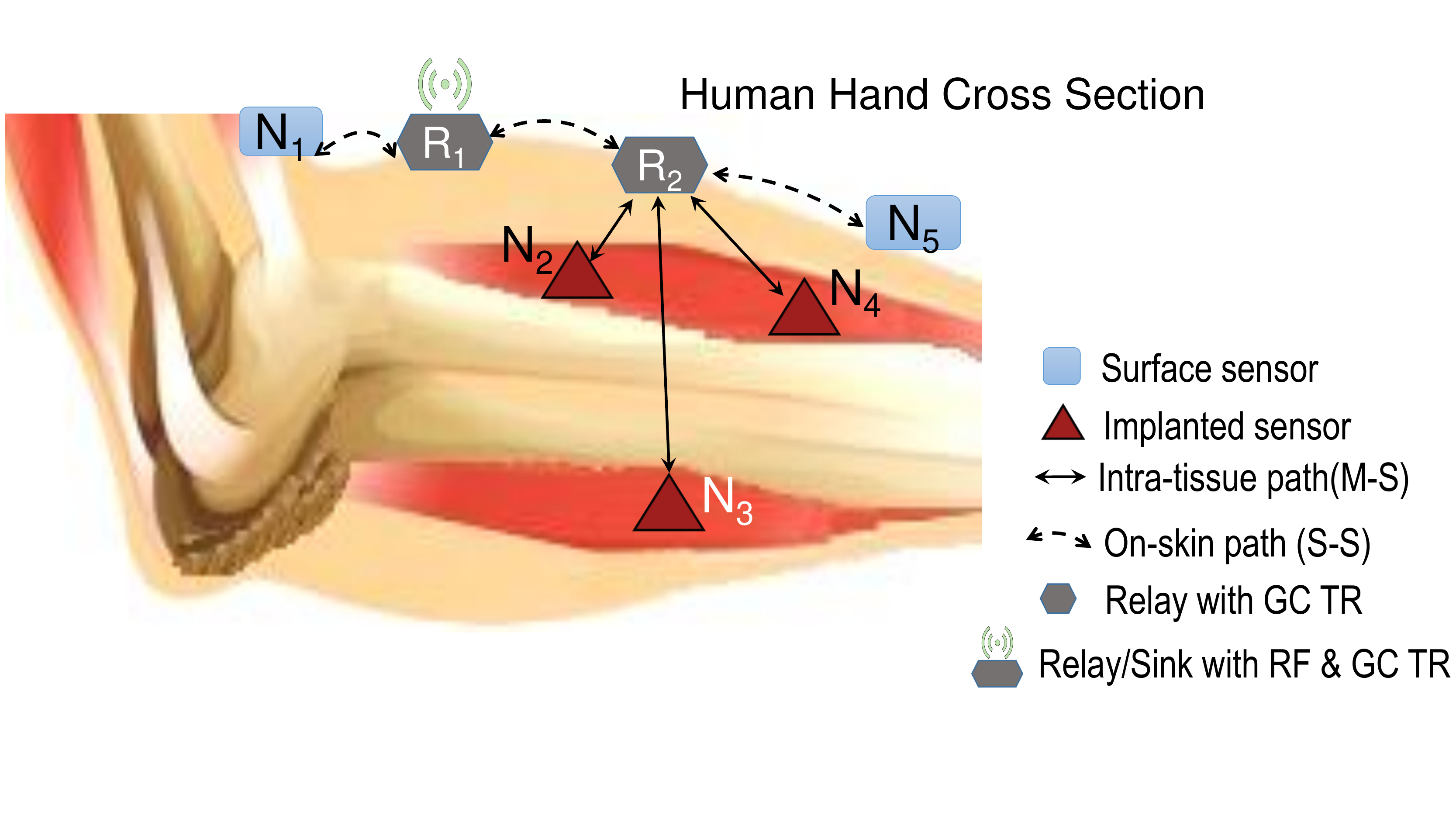}
\vspace{-4mm}
 \caption{Human fore-arm GC-IBN}
 \vspace{-6mm}
 \label{fig:intro}
  \end{figure}

The advent of miniaturized sensing hardware~\cite{MC10} and the possibility of in-situ monitoring of the human body is poised to revolutionize healthcare~\cite{arrayelectrodes}. This vision requires a connected network that will not only report back sensed physiological data, but may also control actuation systems, for e.g., instantaneous drug delivery or electrical discharge for pre-emptive seizure mitigation. Fig.~\ref{fig:intro} shows a sample scenario for a human fore-arm, where the surface nodes $N_1$ and $N_5$ are on-skin sensors while $N_2, N_3$ and $N_4$ are implanted sensors/actuators. Such an intra-body network (IBN) must offer sufficient data transmission rates for timely diagnosis of critical ailments but must also be highly energy conserving, given the practical difficulty in retrieving implants from the body. 

IBN technologies that use over-the-air techniques to establish wireless links, such as capacitive coupling or radio frequency (RF) signaling~\cite{RF}, require a common ground connection that is not possible within implants or incur high attenuation, respectively. As an energy-efficient and safe alternative, we adopt galvanic coupling (GC) based IBN (so called GC-IBN). In this technology, a pair of electrodes in a given node couple a weak modulated signal ($\approx 1\,\mathrm{mW}$) into the tissue obeying the safety limits~\cite{ICNIRP}. Majority of the coupled current passes through the return path of transmitter and a minor part ($\approx$ $-8\,\mathrm{dB}$ for surface node and $\approx$ $-5\,\mathrm{dB}$ for implants) propagates through the tissues. The difference in potential is detected by the electrode pair at the receiver-end that demodulates the signal and extracts the data.

\noindent $\bullet$ \textbf{Problem Definition.} Given the weak signal strength, the network performance of the GC-IBN depends on the length of the links, which is in turn determined by the position of the signal aggregation points or relays. While the GC channel has been modeled earlier in~\cite{tbiocas}, there is no prior work on designing a practical network based on the channel observations. To address this gap, this paper develops a theoretical framework for designing a clustered network, where multiple implants\footnote{interchangeably referred to as a `node' or `sensor'} at various depths inside body tissues ($N_2,..,N_4$ in Fig.~\ref{fig:intro}) are served by an external on-skin relay ($R_2$). The on-skin relays may also forward data to reach a specific relay ($R_1$) that has RF capability to connect with the outside world.

The clustering problem presented here is different from the extensive work for classical wireless sensors that exists today in the following ways: (i) In an IBN, considering the sensor separation in terms of distance alone is not sufficient, but also the specific tissue conduction properties and their relative dimensions must be taken into account (ii) For RF signals, we only need to avoid concurrent signal reception from two or more sources at the receiver that results in detection errors. Instead in the GC-IBN we need to prevent constructive signal combination at any intermediate point that goes beyond permissible limits for tissue safety. (iii) The relative depths at which the implants are located must be considered in clustering to ensure that the harder-to-reach implants have proportionally longer lifetime. (iv) There is no redundancy among nodes within the IBN, and hence, every node is important, and vital to the application. (v) The links between the implants and the external relays exhibit highly asymmetric behavior between uplink and downlink directions. %The M-S path and S-M path can be handled separately with different gain and data rates as we weigh the links based on the channel gain and required bandwidth. But we include only M-S path as we are interested only in implant to relay power consumption. The analysis can also be extended to S-M path if required}
(vi) Finally, there are varying traffic needs between different implants served by the same relay, which may necessitate proximity considerations for certain nodes. 

\noindent $\bullet$ \textbf{Contributions.} The main contributions of this work are:

\noindent 1. We propose the first theoretical clustering framework that provides clear guidelines for placing on-skin relay nodes for embedded implants, while considering the heterogeneous composition and 3-D characteristics of human tissue channels.  %The resulting clusters are adaptable in size, quantity and position.

\noindent 2. We derive acceptable GC-IBN link length distribution that minimizes transmission power for all the nodes, as well as the number of relays, while meeting application demands. This ensures a balanced energy consumption among all the nodes. 

%\noindent 3) We optimize the relay position as a part of the clustering towards energy efficiency and residual energy balance to avoid multiple invasive surgeries.

\noindent 3. We evaluate the effectiveness of the proposed clustering framework using detailed simulation models and experimental studies involving porcine tissue. %The resulting implant life is extended upto several years and also balanced.

The rest of this paper is organized as follows: Sec.~\ref{bg} summarizes the related contributions. Sec.~\ref{sec:model} introduces our GC-IBN system model. Sec.~\ref{sec:cluster} explains our 2-phase clustering framework, which is then analyzed and evaluated in Sec.\ref{sec:eval}. Finally, Sec.~\ref{sec:concl} concludes the paper.

\section{Background and Motivation} \label{bg}
The comparatively short links in GC-IBN ($\approx 30\,\mathrm{cm}$~\cite{tbiocas}) and the varying body channels require dynamic cluster formation. However, analyzing all possible solutions for relay placement is an NP-hard problem, and the short time-scales suggest the use of heuristic approaches. We further discuss additional design considerations for the clustering problem.

\noindent \textbf{Clustering constraints:} Equitable distribution of energy within classical WSNs is achieved by rotating the role of the cluster head (analogous to relay node, in our case)~\cite{CHlife}. The GC-IBN is constrained to have the relay on the skin-surface, and hence the implanted nodes are no longer candidates for role switching. In the general case, WSN protocols assume the bulk of traffic flows in a single direction (i.e., transmit-only sensors and receive-only sink), while a typical GC-IBN with sensors and actuators involve bidirectional traffic. Moreover, as the GC-IBN comprises of non-redundant implants, the network is considered operational until the first implant runs out of energy. This is in contrast to the WSN scenario, where the cluster remains useful as long as a reduced subset of sensors is available. 

\noindent \textbf{3-D propagation:}  Traditional 3D clustering approaches like~\cite{monte} handle all three dimensions equally. However, with the GC transmitter on surface, a receiver at tissue depth (eg., $R_2$ \& $N_3$ in Fig.~\ref{fig:intro}) receives a stronger signal than a receiver on the skin surface at the same distance (eg., $R_2$ \& $N_5$ in Fig.~\ref{fig:intro}) owing to the superior conducting properties of the inner tissues. Straightforward application of techniques, such as K-Means clustering that have been applied to terrestrial WSNs, do not account for the different propagation paths. %Morover  number of clusters ($K$) and initial seed value are unknown. 

\noindent \textbf{Relay positioning constrains:} Classical WSNs have uniform distribution of nodes, which also results in spread of cluster heads throughout the area under study~\cite{multihop}. However, implant locations are influenced by medical applications, and these may result in small pockets of deployment. Thus, the distribution of the relay points in this case is non-uniform. Moreover, relays must forward information among themselves, serving as a conduit for messages among the sensors, instead of direct communication between multiple implant pairs (e.g., $R_1$ and $R_2$ forward information, instead of $N_1$ and $N_2$ directly). This ensures lower energy consumption for the implants, but imposes constraints on the number of nodes connected to relays. Finally, earlier works on relay positioning for on-surface nodes are not suitable for implants~\cite{wcnc}, which makes the current problem scenario novel.

\section{GC-IBN System Model \& Bounds} \label{sec:model}
One of the relay nodes function as the data sink, with RF wireless connectivity to transfer the obtained data to an external monitor ($R_1$ in Fig.\ref{fig:intro}). The other relays, apart from connecting the implants, forward the aggregated and partially processed data towards the sink, forming a two-tier hierarchical architecture. We limit this work to optimize the intra-cluster topology that includes choosing the nodes participating in each cluster and estimating the relay position. We provide an overview of the clustering goals in this section, with the variables listed in Table I. 

We assume a set $\{N_1,..,N_n\}$ of iid nodes embedded as implants in the body, or placed on the body surface, as shown in Fig.\ref{fig:intro}. The position of a node $N_m$ is represented by $L_m\text{=}\{(x_m,y_m,z_m),T_m\}$, where $T_m\in\{skin(S),muscle(M)\}$ is the tissue where $N_m$ in present, and $\{x_m,y_m,z_m\}$ represents the three dimensional coordinates. Specifically, $z_m \in \{0,..,D_{i}\}$ denotes the depth at which the node $N_m$ is present in the tissue $T_m$. $D_i\,\forall i\in T$ is the thickness of chosen tissue layer. Note that we limit this work to skin and muscle tissues, for the purpose of ease of explanation and given that implants are not generally embedded in fat or bone, but the steps can easily be extended to include fat or bone tissues.

\begin{figure}[t]
    \centering
\includegraphics[width=8.6cm,height=3.8cm]{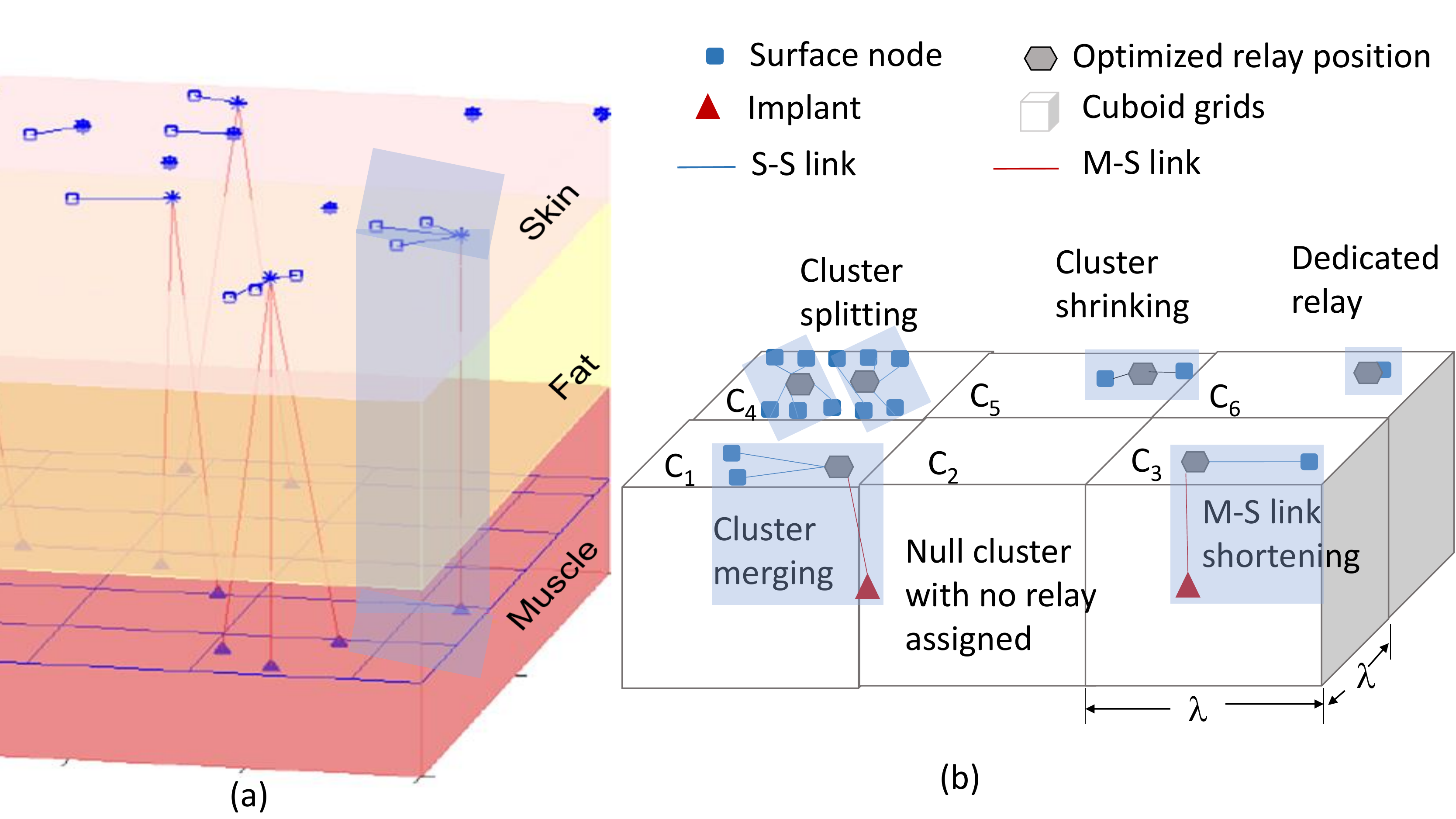}
 \vspace{-2mm}
\caption{(a) Clustered GC-IBN (b) Clustering objectives (gray lines represent uniform grids and shaded blue area denotes optimized clusters)}
\label{fig:2nd}
\vspace{-3mm}
\end{figure}

The primary goal of clustering is to place the relay closer to implants (for short links) as well as to connect more nodes to the relay, even from neighboring clusters (cluster merging) so as to reduce the number of relays required (eg., as illustrated in clusters $C_1,\,C_3\,\&\,C_5$ in Fig.\ref{fig:2nd}(b)). If the number of nodes exceeds the relay capacity (defined below), an additional relay is assigned to handle the overload situation (refer $C_4$ in Fig.\ref{fig:2nd}(b)). Intuitively, no relays are assigned to regions with no nodes (refer $C_2$ in Fig.\ref{fig:2nd}(b)). An isolated node that cannot be reached from existing relays should be allocated to a dedicated relay (similar to $C_6$ in Fig.\ref{fig:2nd}(b)).

The number of nodes in a cluster $C_k,~\forall k \in \{1,..,K\}$ is denoted as $|C_k|\leq n$, where $|.|$ denotes cardinality. $K$ is number of clusters that contains $I_k\leq |C_k|$ implants and $|C_k|\text{-}I_k$ surface nodes. The relay $R_k$ assigned for cluster $C_k$ is on the skin surface at $L_{R_k}$, depth ($z_k\text{=}0$) and is reachable from all the $|C_k|$ nodes through single-hop transmission. Fig.\ref{fig:2nd} shows the scenario where the GC-IBN clusters extend over multiple tissues. 
%The node data rate required is specified as $\eta_m \textgreater 0,\,\forall m\in\{1,..,n\}$. 
We avoid representing the length of the link ($\Lambda_{mR_k}$) between the implant $N_m$ and relay $R_k$ purely in terms of Euclidean distance, considering the presence of heterogeneous tissues between the surface relay and the muscle implant (for surface nodes, the Euclidean assumption still holds). Instead, we approximate the length of the link to be homogeneously co-planar in muscle with the relay assumed to be vertically below the surface containing $L_{R_k}$, and on the plane of the implant in the muscle at $L_{R_k}'$.
\begin{equation}\label{dist}
{\bar{\Lambda}}^2_k\text{=}\begin{cases} \Lambda^2_{iR_k}\text{=}X^2\text{+}Y^2\text{+}Z^2, & T_i\text{=}\{S\}, i\text{=}\{1,..,|C_k|\text{-}I_k\}\\
\Lambda^2_{jR_k'}\text{=}X'^2\text{+}Y'^2\text{+}Z^2, & T_j\text{=}\{M\}, j\text{=}\{1,..,I_k\}
\end{cases}
\end{equation}
where $\bar{\Lambda}_k\text{=}\{\Lambda_{1R_k},..,\Lambda_{nR_k}\},\, X\text{=} |x_i\text{-}x_{R_k}|$, $Y\text{=}|y_i\text{-}y_{R_k}|$, $X'\text{=} |x_j\text{-}x_{R_k'}|$, $Y'\text{=}|y_j\text{-}y_{R_k'}|$, $Z\text{=}z_{\{i,j\}}$ and $\{x_{R_k'},y_{R_k'}\}\in L_{R_k}'$. 
\begin{table}[b]
\vspace{-5mm}
\centering
\caption{\label{tab:overview} Variable definitions and ranges}
\vspace{-3mm}
\footnotesize
\begin{tabular}{l|l}
\toprule
Variable& Definition\\
\midrule
$n$& Total number of nodes \\
$L_m$ & Position $\{(x_m,y_m,z_m),T_m\}$ of node $N_m\in\{1,..,n\}$\\
$T$ & Set of tissues, i.e., $T=\{skin(S)\,,muscle(M)\}$\\
$D(i)$& Thickness of tissue $i$, $\forall i \in T$\\
$z$ & Depth in tissue i.e., $z= \{0,..,D_{\{i,j\}}\},\,i,j\in\{S,M\}$ \\
$\eta_m$ & Required data rate for node $m$, $\forall m\in\{1,..,N\}$\\
$K$ & Quantity of cluster \& relay in GC-IBN\\
$C_k$ & Quantity of nodes in cluster $k$, $\forall k\in\{1,..,K\}$\\
$I_k$ & Quantity of implants in cluster $k$ with  
$C\text{-}I$ surface nodes\\
$L_{R_k}$ & 3D position of relay in cluster $k$, $\forall k\in\{1,..,K\}$ \\
$\Lambda_{mR_k}$ & Transmitter (node) - receiver (relay) link length\\
$Pt_m$ & Transmit power consumed in node $m$, $\forall m\in\{1,..,N\}$ \\
$g^{mR_k}$ & Channel gain through path ${mR_k}$\\ 
$\delta^{mR_k}$ & SNR in path ${mR_k}$\\
%N$_o^{mR_k}$ & Gaussian distributed noise P.S.D in path ${mR_k}$\\ 
%$ f_m$ & Receiver bandwidth\\
$w$ & Link weights based on $\eta$ and $T$\\
NL& List of nodes not yet clustered\\
$\alpha$ & Energy prioritizing factor\\
$\hat{U}$ & $Pt$ uniformity factor\\
$\lambda$ & Length and width of cuboid grid\\
\bottomrule
\end{tabular}
\end{table}

The channel gain between the node and the relay it is connected to can be estimated in terms of the link length as:
\begin{equation}
\bar{g}=\begin{cases}
g_{iR_k}=f_{S-S}(\Lambda_{iR_k}), & i=\{1,..,|C_k|\text{-}I_k\} \\
g_{jR_k}=f_{M-S}(\Lambda_{jR_k'},z_j), & j=\{1,..,I_k\},
\end{cases}
\end{equation}
where $\bar{g}$ is the channel gain vector corresponding to $\bar{\Lambda}_k$. For a surface node to the relay, that we term as skin to skin (S-S) scenario, let $f_{S-S}$ be the function mapping the Euclidean link length to the channel gain between them using the circuit based channel model built with the tissue electrical properties. These two nodes can be the on-skin sensor and the relay (eg., $N_5$ \& $R_2$ in Fig.\ref{fig:intro}), or between two relays (eg., $R_1$ \& $R_2$ in Fig.\ref{fig:intro}). 
Similarly, $f_{M-S}$ maps the approximated implant-relay link length to the multi-tissue channel gain, given the path traverses muscle to skin (M-S). We achieve this by adding the skin and fat impedance to that of the muscle at $L_{R_k}'$ to obtain the heterogeneous M-S path gain. Functions $f_{S-S}\, \&\,f_{M-S}$ have been earlier developed in~\cite{tbiocas}, which we use in this work. In Fig.\ref{fig:intro}, $N_5$ is connected to $R_2$ through the S-S path, while the implants $N_2$, $N_3$ and $N_4$ communicate with $R_2$ via the M-S path. 

%After the IBN is deployed, the channel gain $g'_{mR_k}$ is estimated from the channel SNR ($\delta_{mR_k}$) obtained using received signal strength (RSS) as
%\begin{equation}
%g'_{mR_k}=\frac{\delta_{mR_k}(RSS)N_o^{mR_k}f_m}{Pt_m},\,\forall\, m \in \{1,..,C_k\}
%\end{equation}
% \textbf{XXX-isn't this centralized allocation... why is post deployment needed? How is this used?}.  

%$\Lambda_{mR_k}$ plays a critical role in determining the path loss, $Pt_m$ and the network performance that in-turn is determined by the position, size and quantity of the cluster and hence demands optimization. With this objective, we derive the required bounds that are to be satisfied in GC-IBN. We ignore non-channel dependent parameters (such as sensing, actuation costs) that also also ultimately impact the energy budget of the implant.

As the first step towards our proposed heuristic clustering scheme, we establish the upper and lower bounds on the transmit power that are feasible, as this directly impacts the separation distance.

\noindent $\bullet$ \textit{Lower bound on Pt:} 
The bit error rate must remain below the application demands. This can be achieved by ensuring the channel SNR remains above the desired SNR ($\delta_{mR_k}$) by controlling the minimum required transmit power $Pt_m^{min}$ as:
\begin{equation}\label{eqn:Ptmin}
Pt^{min}_{m}\text{=}\begin{cases}
\frac{{\delta}_{mR_k}N_o^{mR_k}. f_m}{g_{mR_k}}& \forall\, m \in \{1,..,|C_k|\},\Lambda_{mR_k}\textgreater 0 \\
0 & \forall\, m \in \{1,..,|C_k|\}, \Lambda_{mR_k}\text{=} 0,
\end{cases}
\end{equation}
where $N_o^{mR_k}$ is the Gaussian noise P.S.D in $N_m\text{-} R_k$ path with zero mean, % and variance $\varphi$, 
and $f_m$ is the receiver bandwidth. The condition $\Lambda_{mR_k}\text{=}0$ is possible only with surface nodes that also acts as a relay. We ignore this condition for further analysis.

\noindent $\bullet$ \textit{Upper bound on Pt:} $Pt_{m}$ is bounded above by two factors. First, to ensure tissue safety, the maximum transmit power, $Pt_m^{max}$ must satisfy the following condition assuming a single transmission occurs at a time.
\begin{equation}\label{eqn:Ptmax}
Pt_m \leq Pt_{s} \,\forall\, m \in \{1,..,|C_k|\},
\end{equation}
where $Pt_{s}$ is the maximum safe power that can be transferred though the tissues~\cite{ICNIRP}. Second, the lifetime of the implant must be sufficiently long. Assuming M-PSK modulation, the energy consumed by $N_m$ over a period $H_m$ can be estimated using the link budget calculation as,
\begin{equation}\label{eqn:EH}
E^H_m=\frac{E_b^m\,\eta_m\,H_m}{ f_m. log_2 M'_m}  ,\,\forall\, m \in \{1,..,|C_k|\},
\end{equation}
where $E_b^m$ is the bit transmission energy, $M'$ is the modulation level and $\eta_m$ is the data rate required in $N_m$. For lower values of $E_b^m$, the total energy consumption over $H_m$ will also be lower, indicating the possibility of extended $H_m$. With an initial energy store of $E_0$ for the battery life to extend beyond $H_m$, the following condition should be satisfied: 
\begin{equation} \label{cond:HPwithE}
H_m.Pt_{m} \leq E_0,\,\forall\, m \in \{1,..,|C_k|\},
\end{equation}
Using (\ref{eqn:Ptmax}), (\ref{eqn:EH}) and (\ref{cond:HPwithE}), the upper bound on $Pt_{m}$ for the given $H_m$ and $Pt_s$ can now be obtained as:
\begin{equation}\label{eqn:Ptmax2}
Pt^{max}_{m}\text{=}min\{Pt_{s},\frac{E_0}{H_m} \},\,\forall\, m \in \{1,..,|C_k|\}
\end{equation}

\noindent $\bullet$ \textit{Bounds on $\Lambda$:} The maximum link distance that offers the desired node lifetime, SNR and BER, without exceeding $Pt_s$ is the threshold length, $\Lambda^{th}_m$ of node $N_m$, and estimated as:
\begin{equation}\label{eqn:new_th}
\Lambda^{th}_m\text{=}\begin{cases} 
f_{S\text{-}S}'(Pt_m^{max},g_{mR_k})& \forall m \in \{1,..,|C_k|\text{-}I_k\} \\
f_{M\text{-}S}'(Pt_m^{max},g_{mR_k})& \forall m \in \{1,..,I_k\}
\end{cases}
\end{equation}
where, $f'_{S-S}$ \& $f'_{M-S}$ are the inverse functions of $f_{S-S}$ \& $f_{M-S}$ along the S-S and M-S paths respectively, providing the length of the link that offers the gain $g_{mR_k}$ for $Pt_m^{max}$. We assume the threshold link length along S-S path, $\Lambda^{th}_{S-S}$%\triangleq min\{\Lambda^{th}_1,..,\Lambda^{th}_{|C_k|-I_k}\}$
within the cluster as constant as the surface nodes need not be that constrained in terms of energy replenishment. However, the threshold link length of the M-S path, $\Lambda_{M-S}^{th}\triangleq min\{\Lambda^{th}_1,..,\Lambda^{th}_{I_k}\}$ concerning the implants of a cluster vary in $(0,f'_{M-S}(Pt_s,g)]$ as implants with long life require shorter links and vice-verse. 
Hence both singlehop and reliable communication from $N_m$ to $R_k$ is feasible if
\begin{equation}\label{condn:thres-dist}
0\,\textless\, \Lambda_{mR_k}\,\leq \,\Lambda^{th}_m,\, \forall m \in \{1,..,|C_k|\},\,\forall k \in\{1,..,K\}.
\end{equation}

\noindent $\bullet$ \textit{Cluster capacity limit:} \label{cluster_capactiy}
The overall bandwidth requirement of a cluster cannot exceed the capacity ($Q_o$) of its outgoing link that connects $R_k$ to the next hop surface relay or sink. %inter-cluster network (using Min-Cut theorem)~\cite{tbiocas}. 
Hence, $C_k$ is restricted to the set of nodes with the sum of required bandwidth less than $Q_o$. This bounds $C_k$ as follows:
\begin{equation}\label{eqn:clustersize}
1 \leq |C_k| \leq \text{max}\,_{|C_k|} \left(\sum_{m=1}^{|C_k|} \eta_m \leq Q_o\right)
\end{equation}
%where the lower bound $1$ indicates the possibility of single node cluster as explained in Sec.\ref{single_node}.
\subsection{Heterogeneity factors in GC-IBN}\label{hetero}
Given the choice of tissues where the implants are present, the path-loss and energy cost differ. In addition, the bandwidth requirements of the nodes are also not uniform (eg., sensors may require higher uplink bandwidth while actuator may need higher downlink bandwidth). To address this scenario, we capture the various heterogeneity factors in a single-valued \textit{weight} metric.

\noindent $\bullet$ \textit{Heterogeneity from embedded tissue ($T$):} Surface nodes ($T_m\text{=}\{S\}$) incur low energy-conservation costs ($C(S)$) as they are on the skin, hence easily accessible. Implants ($T_m\text{=}\{M\}$) incur a higher cost per unit of energy spent ($C(I)$) as they require invasive procedures for energy replenishment. For extended cluster lifetimes, the overall energy consumption of implants is to be minimized as $C(I) \textless\textless C(S)$ or
\begin{equation}\label{eqn:EIlessEH}
\frac{1}{I_k}\sum_{i=1}^{I_k} Pt_{i} \textless\textless \frac{1}{(|C_k|-I_k)}\sum_{s=1}^{|C_k|\text{-}I_k} Pt_{s},
\end{equation}
for $I_k$ implants and $|C_k|\text{-}I_k$ surface nodes. Hence $\Lambda_{mR_k}$ is to be weighed based on $(T_m,\, z_m)$ to ensure optimal clustering.

\noindent $\bullet$ \textit{Heterogeneity from data rate ($\eta$):} The difference in $\eta$ among the cluster nodes suggests that nodes with higher data rates require longer duty cycles and consume more energy. For instance, assuming $\{N_A$ \& $N_B\in C_k\}$ with $\Lambda_{AR_k}\text{=}\Lambda_{BR_k}$ and $\eta_A\, \&\, \eta_B$ as respective required data rates,
\begin{equation} \label{condn:eta}
\text{if } \eta_A \propto \gamma\, \eta_B,\text{ then } Pt_{A} \propto\gamma Pt_{B}
\end{equation}
for some constant $\gamma \textgreater 1$. This results in an undesirable variation in $Pt_{m}$, $\forall m\in \{1,..,I_k\}$. To ensure equitable energy costs throughout the network with varying $\eta$, the on-skin relay should be ideally moved closer to the node with higher $\eta$ (i.e., closer to node $A$ in (\ref{condn:eta})) to compensate the additional $Pt$ ($Pt_{A}$) required. This can be achieved by weighing the links proportionally with respect to $\eta$.
\subsection{Node weights Estimation}
The critical heterogeneity factors $T$, $z$ and $\eta$ are integrated into an effective weighted link metric $w_m$ for each node estimated as,
\begin{equation}\label{eqn:wt}
w_m=\alpha^{(T_m+z_m){-}1}\displaystyle \frac{\eta_m}{\sum_{i=1}^{|C_k|} \eta_{i}},\forall m \in \{1,..,|C_k|\}
\end{equation}
$\forall k\in\{1,..,K\}$, where the first term from the right weighs the nodes according to the tissues and depths, while second term modifies the weights based on the normalized data rate. $\alpha\in [1,10]$ is the energy prioritizing factor chosen based on the difference desired between $C(I)$ and $C(S)$. {Enumerating} $T_m\text{=}\{S,M\}$ as $T_m\text{=}\{1,2\}$, $\alpha\text{=}1$ denotes $C(I)\text{=}C(S)$ that is suitable for setting up a short term GC-IBN, with equal life span for surface nodes and implants. On the other hand, $\alpha\text{=}10$ sets much higher C(I) suitable for long term deployments. %The link length $\Lambda_{mR_k}$ can be 
%\begin{equation} \label{eqn:wtlambda}
%w\Lambda_{mR_k}=w_m \Lambda_{mR_k}
%\end{equation} 
\subsection{The Uniformity factor}
As one of the design goals, we target proportional power consumption among all implants. To quantify this concept, we devise a quality metric $\hat{U}$ that determines an approximate percentage of residual energy in the remaining nodes when the first implant is lost.
%However, if all the implants in a cluster have less than $10$ or $20\%$ of residual energy when the first implant is drained out, then, they all can be replaced at once. With this perspective, we devise a quality metric to control the proportional power consumption among the implants as
\begin{equation}\label{eqn:UB}
\frac{Min(\Lambda_{mR_k})}{Max(\Lambda_{mR_k})} \textgreater \hat{U} ,\,\forall m \in \{1,..,I_k\},
\end{equation}
%where $\hat{U}$ is the desired uniformity. 

For example, $\hat{U}\text{=} 0.8$ ensures that when $N_x\triangleq max_x (Pt_{x})$ gets depleted with $Pt_{x}.H_x=E_0$, the residual energy in other live implants will be $\leq$ $20\%$ of $E_0$. Accordingly, in our clustering framework, we restrict $C_k$ to include only those implants that satisfy (\ref{eqn:UB}) for an estimated relay position. The number of nodes satisfying the uniformity constraint might be increased if $\hat{U}$ is relaxed to a lower value (e.g. from $0.8$ to $0.7$). The influence of $\hat{U}$ on $K$ is analyzed in Sec.\ref{sec:eval}(S5). 

 \vspace{-1mm}
\section{GC-IBN Cluster Topology Design}\label{sec:cluster}
In this section, we propose our heuristic two-phase iterative clustering framework for GC-IBN. In the first phase or the Initial Cluster Approximation Phase (ICAP), we generate the initial approximation of $K$ and $C_k$ using the node positions. With no prior knowledge on $K$ or $L_{R_k}$, we use neighborhood learning that uses lower energy than conventional message passing techniques. In the second phase, we perform the Nearest Neighbor based Iterative Cluster Optimization (NICO) until the achieved clustering is deemed optimal. Thus, through this approach, the quantity, size and position of the clusters are iteratively optimized for energy efficiency under performance constraints. 
\vspace{-1mm}
\subsection{Phase I. Initial Cluster Approximation Phase (ICAP):}
In this phase, we use definitive cuboid grid based clustering as (i) it identifies and connects outliers while no nodes are left under-classified, a common possibility among density based approaches~\cite{monte}, and (ii)partitioning the node space into finite number of cuboids requires fixed number of executions that depends only on the number of cuboids irrespective of $n$. The cuboids are preferred over the spheres as the grid size is different in the third spatial dimension (i.e., height varies). In addition, the cuboid packing avoids area overlaps or gaps so that every node is considered only once for clustering.

\noindent \textit{Grid size estimation:} 
Larger clusters (i.e., with higher $(|C_k|)$) require longer average link lengths, consuming more $Pt$, while smaller clusters increase the number of clusters $K$. In addition, the clusters should also satisfy the link length condition in (\ref{condn:thres-dist}). We use the unit-cuboid grids to identify the clusters of right size while separating the nodes by a maximum threshold distance of $\Lambda_k^{th}\text{=}min(\Lambda^{th}_{S-S},\Lambda_{M-S}^{th})$.

\noindent \textbf{Theorem \ref{sec:cluster}.1}: 
In a unit cuboid of length and width being $\lambda\text{=}\Lambda^{th}_k/\sqrt{2}$, the maximum S-S or M-S link length between the nodes and random relay positions satisfies $\Lambda^{max}_{mR_k} \leq \Lambda^{th}_k$. 

\noindent \textbf{Proof:} The link lengths of nodes in the grid are independent and have the same distribution. Let the maximum link length in $C_k$ be $\Lambda^{max}_{mR_k}$. Using the cumulative distribution function of $\Lambda_{mR_k}$ (formulated in Appendix.1),

$P(\Lambda^{max}_{mR_k}\textgreater\lambda\sqrt{2})$

$=P(\Lambda^{max}_{mR_k}\leq\lambda\sqrt{2}+1)-P(\Lambda^{max}_{mR_k}\leq\lambda\sqrt{2})$ 

$= F_\Lambda(\lambda\sqrt{2}+1)^n-F_\Lambda(\lambda\sqrt{2})^n = 0 \hfill\blacksquare$

\noindent The grid size defined with $\lambda\text{=}min(\Lambda^{th}_{S-S},\Lambda_{M-S}^{th})$  satisfies (\ref{condn:thres-dist}). Hence, we use $\lambda$ for partitioning with splitting index as 
\begin{equation} \label{gridpts}
\{x_\lambda,y_\lambda,z_\lambda\}=\{X_1\text{+}a\lambda,\,Y_1\text{+}b\lambda,\,D\},
\end{equation}
where $a\text{=}\{0,1,..,\lceil (X_2\text{-}X_1)/\lambda\rceil \}$, $b\text{=}\{0,1,..,\lceil (Y_2\text{-}Y_1)/\lambda\rceil \}$, $\lceil.\rceil$ is the ceil function, $D$ is the tissue thickness comprising of skin, fat and muscle, forming the third dimension and $\{X_1,Y_1\}$ \& $\{X_2,Y_2\}$ are the surface dimensions of the body area of interest. Prior clustering, the $n$ nodes in given body area are included in the `not-clustered list', NL. The three dimensional grid virtually partitions the given tissue into $K\text{=}(|a|-1)(|b|-1)$ cuboids or clusters (refer Fig.\ref{fig:2nd}(b)). The set of $C_k=\{N_i,\,\forall i\in \{1,..,|C_k|\}:\,|C_k| \leq n\}$ nodes enclosed by a cuboid $k$ participates in the cluster $k\,\forall k\in\{1,..,K\}$. %, i.e.,
%\begin{equation}
%C_k\text{=}\{N_i: X_1\text{+}a\leq x_i\leq X_1\text{+}a\lambda; Y_1\text{+}b\leq y_i\leq Y_1\text{+}b\lambda\}
%\end{equation}
%$\forall i\in [1,n]$. 
The NL is then updated to remove the clustered nodes as $NL=NL\backslash C_k,\forall k\in\{1,..,K\}$. The so formed grids give an initial approximation on the number of clusters $K$ in the given area. %The PDF of $\Lambda_{mR_k}$ in cluster $k$ is plotted in Fig.\ref{fig:stat_results}(left). 
 
\noindent \textbf{Lemma \ref{sec:cluster}.1}: 
%For $p\leq a$ and $q\leq b$, 
If $C_1\text{=}\frac{\Lambda_{S-S}^{th}}{\sqrt{2}}$, $C_2\text{=}X_2\text{-}X_1$, $C_3\text{=}Y_2\text{-}Y_1$, $p\in [\lceil \frac{C_2}{C_1}\rceil,C_2]$ and $q\in [\lceil \frac{C_3}{C_1}\rceil,C_3]$, the CDF of $K$ is given by $\lceil \frac{p(C_1-1)-(C_2-p)}{p(C_1-1)}\rceil \lceil  \frac{q(C_1-1)-(C_3-q)}{q(C_1-1)}\rceil$.

\noindent \textbf{Proof:} Using (\ref{condn:thres-dist}) and the observation that $\Lambda^{th}_{S-S} \textless \Lambda^{th}_{M-S}$ for the same $Pt$ \cite{tbiocas}, we see that $\lambda$ is uniformly distributed in $(0,\Lambda^{th}_{S-S}/\sqrt{2}]$. Assuming the minimum value of  $\lambda$ as $1$, the CDF of $\lambda$ is given by
\begin{equation} \label{lambdaCDF}
 {F_{\lambda}=\begin{cases} 0, & \lambda \textless 1\\
\frac{\lambda-1}{C_1-1}, & 1 \leq \lambda \leq C_1  \\ 1 & \lambda \textgreater C_1\\
 \end{cases}}
\end{equation}
Let $P=|a|-1$. The CDF of $P$ can be derived using (\ref{gridpts}) as
\begin{equation} \label{aCDF}
 {F_P(p)=F_{\lambda}(\frac{C_2}{p}) =\begin{cases} 0, & p\textless \lceil \frac{C_2}{C_1}\rceil\\
1-\frac{\frac{C_2}{p}-1}{C_1-1},& \lceil \frac{C_2}{C_1}\rceil \leq p \leq C_2\\
 1, & p \textgreater (C_2) \end{cases}}
 \end{equation}
using variable transformation technique. If $Q=|b|-1$, then $P$ and $Q$ share similar distribution and are also independent. The joint CDF of $F_K(P,Q)$ can be obtained from (\ref{aCDF}) as a joint distribution of $P$ and $Q$ as $F_K(p,q)=$
$$\begin{cases} 0, & p\textless \lceil \frac{C_2}{C_1}\rceil, q\textless \lceil \frac{C_3}{C_1}\rceil\\
\lceil 1\text{-}\frac{\frac{C_2}{p}-1}{C_1-1} \rceil \lceil 1\text{-}\frac{\frac{C_3}{q}-1}{C_1-1} \rceil, & \lceil \frac{C_2}{C_1}\rceil \leq p \leq C_2,\lceil \frac{C_3}{C_1}\rceil \leq q \leq C_3  \\
1, & p \textgreater C_2,q\textgreater C_3 \hfill \blacksquare
\end{cases}$$
%For determining the node distribution, we consider the tissue layers as stacked 2D planes as (i) their thicknesses are negligible compared to the transmitter-receiver separations and (ii) distance computation is inversely proportional to channel gain at depths.
%The density distribution of $|C_k|$ is binomial on $\lambda$ \& $n$ given by, $Pr\{|C_k|=c\, |\, n=N\} = \binom N c Q^c (1-Q)^{N-c}$, 
%where $Q\triangleq \frac{\lambda^2}{A}$ %with CDF, $F_Q(q)\text{=}F_{\lambda}(\sqrt{\frac{A}{q}})$ 
%and $A\text{=}(X_2-X_1)(Y_2-Y_1)$ is the body area considered. %The $|C_k|$ distribution is plotted in Fig.\ref{fig:stat_results}(right) for various values of $\lambda$. 
At the end of this phase, NL=$\{\varnothing\}$ as all the nodes are included in some cluster.

\subsection{Phase II. Nearest Neighbor based Iterative Cluster Optimization (NICO)}
\begin{figure}[t]
   \vspace{-1mm}
 \centering
\includegraphics[width=8cm,height=9.5cm]{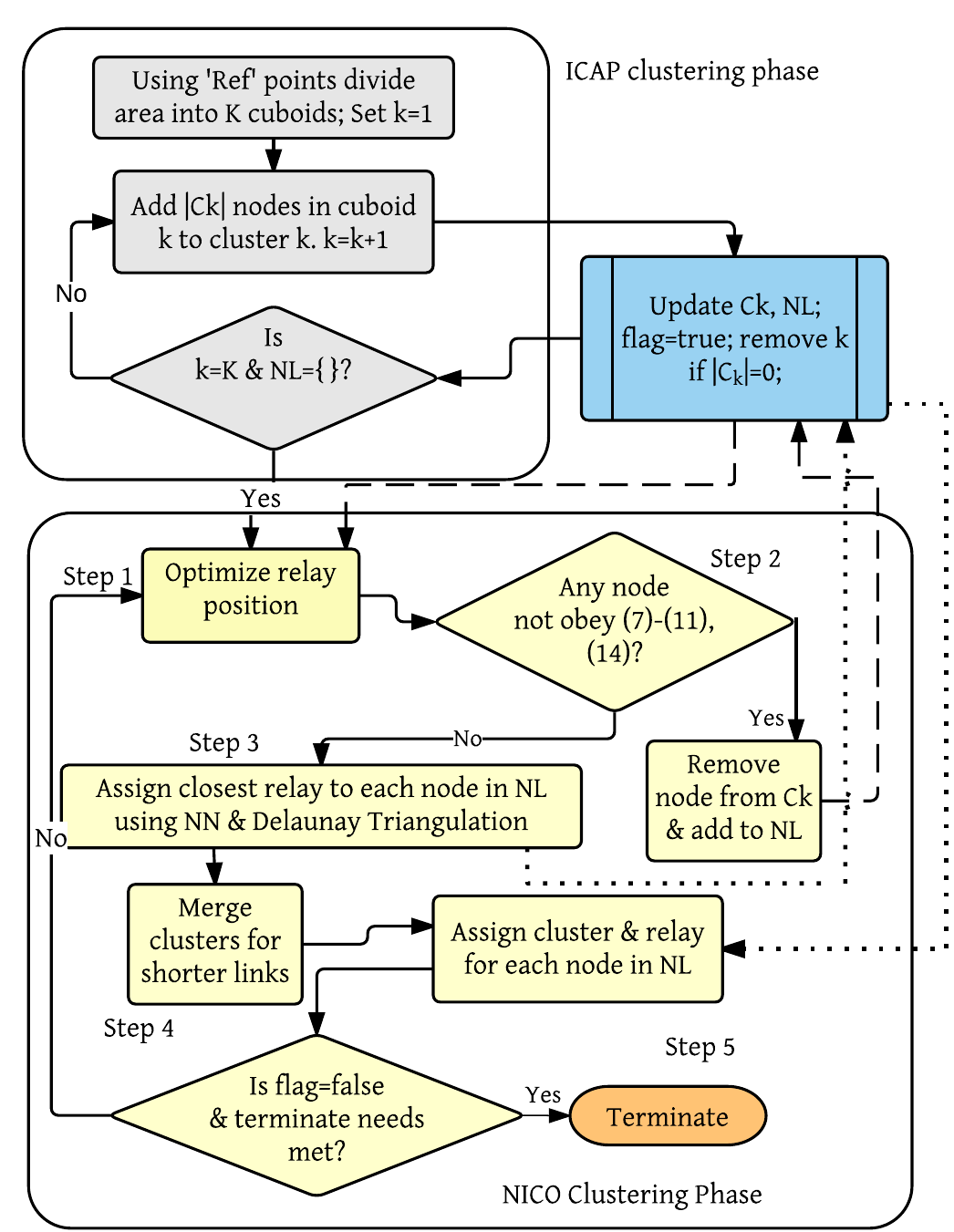}
  \vspace{-1mm}
 \caption{Two Phase Clustering Algorithm}
 \label{fig:clusterflow}
  \vspace{-4mm}
 \end{figure}

In this phase, we semi-locally optimize the clusters from the approximations obtained in ICAP by iteratively adjusting $K$ and $C_k,\,\forall k\in\{1,..,K\}$. The NICO phase comprises of the following five steps that are iterated in sequence. The description below follows the flowchart given in Fig.~\ref{fig:clusterflow}.

\noindent \textbf{Step 1. Relay position optimization ($\hat{L}_{R_k}$):}
The NICO phase starts with identifying the optimal relay position $\hat{L}_{R_k}$ for each $C_k,\,k\in\{1,..,K\}$ towards minimizing $Pt_m,\,\forall m\in \{1,..,|C_k|\}$ and balancing $Pt_m,\,\forall m\in \{1,..,I_k\}$ while achieving the link shortening and cluster shrinking objectives given in Fig.\ref{fig:2nd}(b). For instance in Fig.\ref{fig:intro}, the total link length, and hence, the path loss experienced by the implant $N_2$ can be reduced by bringing $R_2$ closer to $N_2$. A dramatic decline in $Pt$ obtained by reducing the link length by just a few centimeters is described in Sec.\ref{sec:eval}(S1). 
%The optimum position of the relay, $\hat{L}_{R_k}$ that reduces the link length of multiple (or even all) nodes in a cluster is estimated in~\cite{wcnc} using the iteratively re-weighted least squares based modified Weiszfeld algorithm. \textcolor{blue}{XXX- if you are mentioning this here, don't cite WCNC work... also Infocom double blind policy does not allow multiple self citations... just cite once the channel model} This algorithm estimates $\hat{L}_{R_k}$ closer to all the $|C_k|$ nodes in cluster $k$ given by,
The optimal relay position $\hat{L}_{R_k}$ that reduces the link length of multiple (or even all) nodes in a cluster using the weights in (\ref{eqn:wt}) can be estimated as,
\begin{equation}\label{relay_dist}
\hat{L}_{R_k}\text{=}argmin_{L_{R_k}} \sum_{m=1}^{|C_k|} w_m.\Lambda_{mR_k},
\end{equation}
\begin{equation*} {\small
\begin{aligned}
s.t & \,\,(\ref{relay_dist}.i)\,\, \Lambda_{mR_k} \leq \Lambda^{th}_{M-S},\, m\in[1,..,I_k]\\
& \,\, (\ref{relay_dist}.ii)\, \Lambda_{nR_k} \leq \Lambda^{th}_{S-S},\, n\in[1,..,|C_k|-I_k]\\
\end{aligned}}
\end{equation*}
for $I_k$ implants and $|C_k|\text{-}I_k$ surface nodes in $C_k$. Constraints (\ref{relay_dist}.i) \& (\ref{relay_dist}.ii) limits the link length to the threshold link lengths estimated for M-S and S-S paths. Although the estimated $\hat{L}_{R_k}$ significantly improves node life than the conventional relay positions, it tends to overfit by penalizing a few nodes making their $\Lambda_{mR_k}$ longer in the interest of minimizing the sum of link lengths. 
The issue becomes critical when implants are penalized. (\ref{eqn:wt}) weighs the implants heavier to avoid this. However, when there are multiple implants in the cluster ($I_k\textgreater 1$), few of them might be penalized. Hence, the optimization problem in (\ref{relay_dist}) needs modification, so that it extends the battery life of all the nodes, as well as balances the implant energy.

The deviation in energy consumption among the implants is due to the variation in $w\Lambda_{mR_k}, m\in\{1,..,I_k\}$. Aiming only for the balanced residual energy would lead to near-zero deviation in $w\Lambda_{mR_k}$ with much longer links. Hence, for each cluster, we reformulate (\ref{relay_dist}) by (i) using a weighted $L_1$ norm to prevent overfitting, and (ii) by adding a log barrier function to restrict the non-negative constraints as follows:
\begin{equation}\label{prob:3} {\small
\text{min}%_{L_{R_k}} 
\sum_{i\text{=}1}^A \left( w_i\Lambda_{iR_k} \text{+}\, \gamma\, w \parallel L_i \text{-}L_{R_k}\parallel^1_1 -\mu log(p_1\text{+}p_2) \right)}
\end{equation}
\begin{equation*} {\small
\begin{aligned}
s.t & \,\,(\ref{prob:3}.i)\,\,\Lambda^{th}_{M-S}-\Lambda_{mR_k}+p1=0,\, m\in[1,..,I_k]\\
& \,\, (\ref{prob:3}.ii)\, \Lambda^{th}_{S-S}-\Lambda_{nR_k}+p_2=0,\, n\in[1,..,|C_k|-I_k]\\
%& \,\,(\ref{prob:3}.iii)\, u\in [0,1]\\
%& \,\,(\ref{prob:3}.iv)\, v\in [0,1]\\
\end{aligned}}
\end{equation*}
where $\gamma\text{=}(u\text{-}\,1)v$ is the $L_1$ penalty parameter and $\mu log(p_1\text{+}p_2)$ is the barrier function with $p_1,\,p_2$ as the slack variables. The binary variable $u$ becomes $1$ for $I_k\text{=}0$ and becomes $0$ for $I_k\textgreater 0$, i.e., when $I_k\text{=}0$, $L_{R_k}$ relies only on the surface node positions. The binary variable $v$ tailors the problem for $I_k\textgreater 1$ and $I_k\leq 1$ conditions. The variable $A\text{=}u|C_k|\text{+}(1\text{-}u)I_k$ tunes the objective function towards energy efficiency for cluster without implants and towards energy balance in cluster with implants.

%\begin{equation}\label{prob:3} {\small
%\hat{L}_{R_k} \text{=} \text{min}_{L_{R_k}} \sum \left( \sum_{i\text{=}1}^{A} w_i\Lambda_{i\text{-}R_k} \text{+}\, \gamma \boldsymbol{w\Lambda_{R_k}} \right)}
%\end{equation}
%\begin{equation*} {\small
%\begin{aligned}
%s.t & \,\,(\ref{prob:3}.i)\,\, \Lambda_{mR_k} \leq \Lambda^{th}_{M-S},\, m\in[1,..,I_k]\\
%& \,\, (\ref{prob:3}.ii)\, \Lambda_{nR_k} \leq \Lambda^{th}_{S-S},\, n\in[1,..,|C_k|-I_k]\\
%& \,\,(\ref{prob:3}.iii)\, u\in [0,1]\\
%& \,\,(\ref{prob:3}.iv)\, v\in [0,1]\\
%\end{aligned}}
%\end{equation*}
The optimization problem is convex with $L_1,\,L_2$ norms and affine constraints. \textcolor{black}{We solve it using the interior point method that finds the feasible $\hat{L}_{R_k}$ in the descent direction, estimated from the Newton step on the equivalent Karush-–Kuhn-–Tucker equations (obtained via linear approximations)~\cite{interior}. The Interior point method has fast convergence rate with fewer iterations towards precise solution but requires a suitable starting point}. We estimate the initial relay position at the cluster centroid on surface obtained as $\{\frac{1}{|C_k|} \sum_{m=1}^{|C_k|} L_m(x_m), \frac{1}{|C_k|} \sum_{m=1}^{|C_k|} L_m(y_m),0\}$ with $\mu=0.1$. Note that the heterogeneity conditions in (\ref{eqn:EIlessEH}) \& (\ref{condn:eta}) are implicitly handled by $w$ and hence, are not repeated in the optimization problem. 

\noindent \textbf{Step 2. Cluster Reformation:}
In this step we verify if the clusters obtained above satisfy the limits on $Pt$, $\Lambda_{mR_k}$, $|C_k|$ and $\hat{U}$ derived in (\ref{eqn:Ptmax2})%, (\ref{condn:thres-dist}),(\ref{eqn:clustersize}), 
-(\ref{eqn:EIlessEH}) and (\ref{eqn:UB}). The non-conforming nodes are removed from $C_k$ and added back to NL for reclustering as follows. 
In a cluster that does not satisfy  (\ref{eqn:Ptmax2}), (\ref{condn:thres-dist}), (\ref{eqn:EIlessEH}) and (\ref{eqn:UB}), the implant with the longest link length is removed from the cluster and added back to NL as follows.
\begin{equation}\label{lengthsplit}
NL\text{=}\{NL\cup N_m:m\text{=}Max_m\, (\Lambda_{mR_k})\};\, C_k\text{=}\{C_k\backslash N_m\}
\end{equation}
$\forall m\in\{1,..,I_k\}$. If (\ref{eqn:clustersize}) is not satisfied, then the node with $max(\eta_m),\forall m\in\{1,..,|C_k|\}$ is removed from cluster and added back to NL.
\begin{equation}\label{etasplit}
NL\text{=}\{NL\cup N_m:m\text{=}Max_m\, (\eta_m)\};\, C_k\text{=}\{C_k\backslash N_m\}
\end{equation}
 
\begin{figure}[t]
 \centering
\includegraphics[width=8.5cm,height=3.6cm]{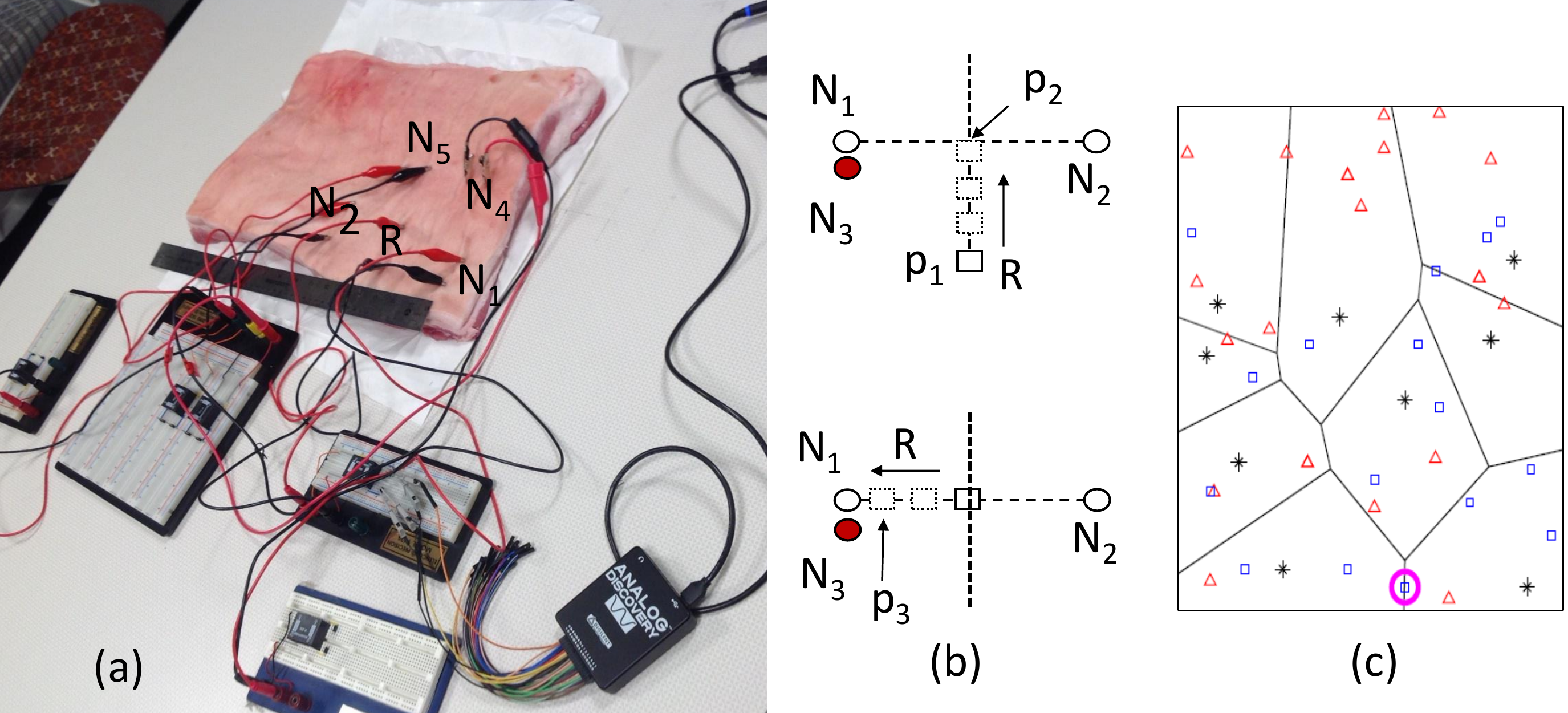}
 \caption{(a) Porcine experimental set-up (b) Relay position changes (c) Voronoi region samples of relay (*) positions; \textcolor{red}{$\triangle$} - implants; \textcolor{blue}{ $\square$} - surface nodes }
 \label{fig:expr_setup}
  \vspace{-5mm}
 \end{figure}

\noindent \textbf{Step 3. Nearest Relay Assignment:} Next, we assign the nodes in NL to the closest cluster. We use a combination of Delaunay Triangulation and Nearest Neighbor algorithms for the purpose. Using the triangulations, the Voronoi region $V_k$ of a relay $R_k, \forall k\in \{ 1,..,K\}$ is determined as the locus of the skin and muscle regions that has $R_k$ as the nearest neighbor.
\begin{equation}
V_k=\{L_m | \Lambda_{mR_k} \leq \Lambda_{mR_j}\}, \forall k,j\in\{1,..,K\},k\neq j
\end{equation}
%Triangulation time complexity - O(nlog n) + O(n^2).
%Delaunay Triangulation is used for obtaining the Voronoi regions as it supports queries such as Nearest Neighbor as well as incremental inclusion and removal of nodes.
Using $V_k$ (refer Fig.\ref{fig:expr_setup}(c)), and the node position as query parameters, our Nearest Neighbor algorithm finds the relay offering shortest $\Lambda_{mR_k}$ and satisfying (\ref{eqn:Ptmax2}), (\ref{condn:thres-dist})-(\ref{eqn:EIlessEH}) \& (\ref{eqn:UB}). This step has a low complexity of $O(Klog(n))$ 
%O(K log K)+ log (n)	
as it avoids the distance computation of every possible node-relay combination. 
When a node is equidistant from multiple relays, (such as the one enclosed in circle in Fig.\ref{fig:expr_setup}(c), we choose the relay with less load to assign the node. %The NN technique is suitable for the GC-IBN like sparse network with the size of network larger than the communication range. 

Note that in this step, the $\Lambda_{mR_k}$ does not rely on the definitive grid dimensions but only on $\Lambda^{th}_{M-S}\, \&\,\Lambda^{th}_{S-S}$. Hence $\Lambda_{mR_k}$ can exceed the cuboid dimension $\lambda$ specified in ICAP. This enables cluster shrinking or expanding just to cover the conforming node positions. Also, $\Lambda_{mR_k}$ distributions for surface nodes differ from that of implants based on $\Lambda^{th}_{S-S}$ and $\Lambda^{th}_{M-S}$, as illustrated later in Fig.\ref{fig:stat_results}.

\noindent \textbf{Step 4. Cluster Reassignment \& Merging:} If the new $\hat{L}_{R_k}$ offers shorter $\Lambda_{mR_k}$ to a node in the neighborhood cluster $C_j$ as given below,
\begin{equation}
\Lambda_{xR_k} \textless \Lambda_{xR_j}:N_x\in C_j,k\neq j,\forall k,j\in\{1,..,K\},
\end{equation}
then \noindent $N_x$ is reassigned to $C_k$, if its inclusion satisfies (\ref{eqn:clustersize}). The link lengths can be compared using the generated  Voronoi regions and Nearest Neighbor algorithm in Step 3. As a consequence of this reassignment in a semi-distributed fashion, if $C_j\text{=}\{\varnothing\}$, then it is deleted with no relay assignment satisfying the null cluster objective in Fig.\ref{fig:2nd}(b). %Thus, in this step, $\Lambda_{mR_k}, K$ and $C_k$ are adapted based on $\hat{L}_{R_k}$.  

\noindent \textbf{Step 5. Dedicated Relays:} \label{single_node} Finally, the nodes in NL that cannot be included in the existing clusters are assigned a dedicated relay, forming clusters with $|C_k|\text{=}1$ and increasing $K$ by the size of NL. This forms the basis of the lower bound in the cluster size defined in (\ref{eqn:clustersize}) that can be merged among themselves or with other clusters in Step.2 of next iteration, if the required conditions are satisfied (see Fig.\ref{fig:2nd}(b)). Any change in the cluster participation from the previous iteration (marked by $flag$ in Fig.\ref{fig:clusterflow}) requirers a new run of the $L_{R_k}$ optimization, followed by the iterative execution of all the steps in NICO phase until the termination criteria (given below) are met.

\noindent \textbf{Termination Criteria:} Fixing the number of clustering iterations can result in many more iterations than actually required. To overcome this, we define our NICO termination criteria as follows. (a) NL=$\{\varnothing\}$, indicating that every node participates in a cluster,(b) $K \leq n$, which is ensured in Step.2, where relays with no node assignments are removed, and (c) $flag$=false, indicating no cluster change in the current iteration.

The time complexity of ICAP is $O(Kn)$ while that of NICO is $O(Kn\,log(n/\epsilon))$+$O(nK)$+$O(Klog\,n)$+$O(Kn)$+$O(n)$ and $O(nlog\,Kn)$ for $NL$ and $flag$ update with an iteration complexity of $O(\sqrt{n}K)$\cite{lp}. Thus the overall worse case complexity of the framework is  $O(n^{\frac{3}{2}}K^2log\, n)$. The resulting $\Lambda_{mR_k}$ and $K$ are substantially reduced from that of the ICAP phase as illustrated in Fig.\ref{fig:stat_results} \& Fig.\ref{fig:cluster2d}. The algorithm is executed offline prior network installation to estimate the relay positions. The NICO phase is repeated periodically after the implants are inserted to 
accommodate channel gain variations that can be identified from received signal strength. 
The required changes on relay position can be made by the care-giver without specialized expertise. 
 
\section{Performance Evaluation}\label{sec:eval}
 \begin{figure}[b]
 \centering
  \vspace{-3mm}
\includegraphics[width=8.5cm]{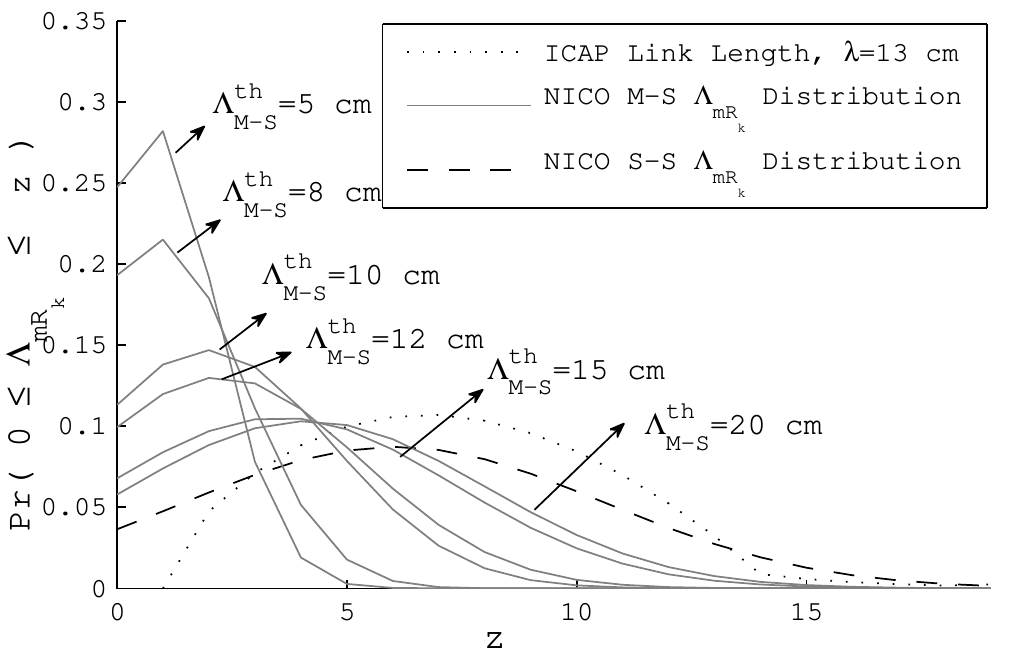}%stat_results1.pdf}
\vspace{-4mm}
 \caption{$\Lambda_{mR_k}$ distribution in ICAP and NICO phases} %(left) and $|C_k|$ distribution (right)}
 %\vspace{-2mm}
 \label{fig:stat_results}
 \end{figure}

In this section, we: (i) analyze the impact of shortening $\Lambda_{mR_k}$ using empirical measurements on porcine tissue (chosen for similarity in properties with human tissues) and verify the simulation parameters with actual measurements in scenarios S1 and S2; (ii) evaluate the proposed GC-IBN clustering and relay positioning framework on optimizing $\Lambda_{mR_k},\,K,\,C_k$ and $Pt_m$ using a galvanic coupled human forearm simulation model that computes the channel gain using the tissue equivalent electrical parameters in a circuit model from~\cite{tbiocas}. 
%\textcolor{blue}{include some additional information in this paper. Also I suggest clearer partitions- separate first the experimental and simulation parts into clearly defined subsections. For the simulation: have separate parts for- all surface nodes, single implant, and then general surface and implant scenarios... you can have further sub-parts for the general scenario with more detailed evaluations}. 
Using the simulation, we compare the optimal link length obtained in the NICO phase with the sub-optimal link length obtained in ICAP phase in scenario S3. We then analyze the power consumption in clusters that have $I_k\text{=}0$ (in S4), $I_k\text{=}1$ (in S5) and $I_k\textgreater 1$ (in S6). Finally in S7, we analyze the clustering efficiency in terms of $K$. For the analyses, the value of $\alpha$ defined in (\ref{eqn:wt}) is assumed to be $4$, unless specified otherwise.

A tissue sample with skin, muscle and fat of dimensions $42\times 25 \times 6 \,\mathrm{cm^3}$, from the porcine shoulder is cleaned and moistened for electrode attachment. The experimental network is composed of a relay ($R$), 2 surface nodes ($N_1$ \& $N_2$ that are $10\,\mathrm{cm}$ apart) and an implant ($N_3$, below $N_1$) by fixing the electrode pair from each node to the tissue as shown in Fig.\ref{fig:expr_setup}(a). We use a multi-channel signal generator and oscilloscope, along with the OEP PT4 1:1 pulse transformers to isolate the transmitter and receiver.

\begin{figure}[t]
 \centering
  %\vspace{-1mm}
\includegraphics[width=8.5cm,height=3.6cm]{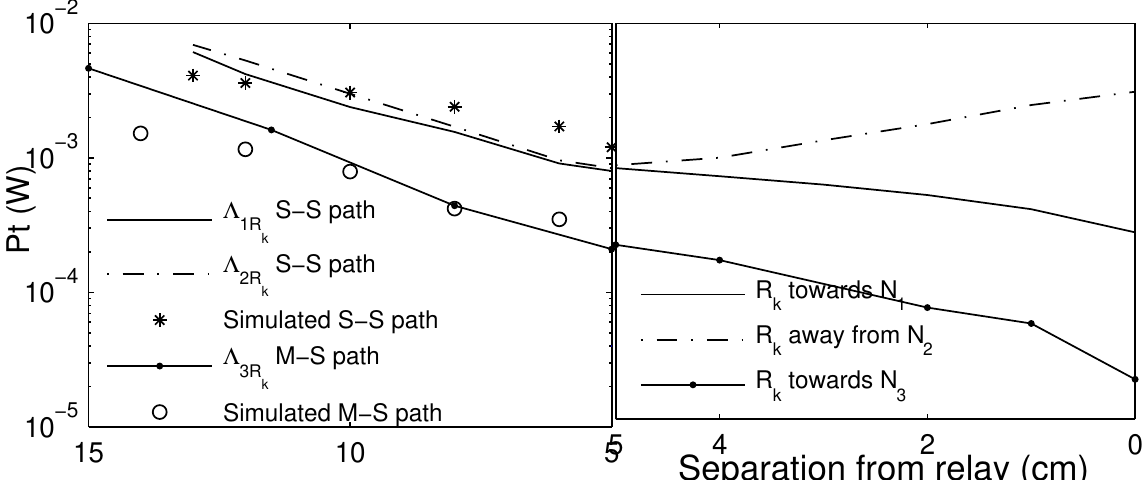}
\vspace{-1mm}
 \caption{Power consumed in $N_1$, $N_2$ and $N_3$ for relay positions in Fig.\ref{fig:expr_setup}.(b)}
   \vspace{-5mm}
 \label{fig:exprplot}
 \end{figure}

\noindent \textbf{S1. Impact of shortened $\Lambda_{mR_k}$ on $Pt$:} Here, we highlight the dramatic reduction in the transmission powers $Pt_1$, $Pt_2$ \& $Pt_3$ for the nodes $N_1$, $N_2$ and $N_3$, respectively, for communicating with $R$ (refer Fig.\ref{fig:expr_setup}(b)(top)) when $R$ is brought closer. 
When $R$ is moved from $p_1$ ($14\,\mathrm{cm}$ from $N_1$ \& $N_2$) to $p_2$ ($5\,\mathrm{cm}$), $Pt_{1}$ and $Pt_{2}$ drops from $6.5\,\mathrm{mW}$ to $0.8\,\mathrm{mW}$ over the S-S path (refer Fig.\ref{fig:exprplot}(left)). Owing to the lower loss in M-S path, bringing $R$ closer to $N_3$ by the same distance substantially cuts down $Pt_{3}$ from $4.6\,\mathrm{mW}$ to $0.2\,\mathrm{mW}$. 

\noindent \textbf{S2. Moving relay closer to implant:}
When $R$ is moved from $p_2$ to $p_3$ (refer Fig.\ref{fig:expr_setup}(b)(bottom)), $Pt_1$ and $Pt_3$ drops even lower to $0.2\,\mathrm{mW}$ and $20\,\mathrm{\mu W}$ respectively, while $N_2$ is penalized by increasing $Pt_2$  from $0.8\,\mathrm{mW}$ to $3\,\mathrm{mW}$ (refer Fig.\ref{fig:exprplot}(right)). With this optimized M-S $\Lambda_{mR_k}$, the life\footnote{By Peukert's law, 
node life=$\left(\frac{\text{Battery capacity} (240\,\mathrm{mAh})}{\text{duty cycle (10\%)}\times\text{Load}}\right)\times\text{External factors}$. The load current is derived from $Pt$ and power consumption from other node functions ($\approx 0.1\,\mathrm{mW}$). %The external factor (discharging rate) is assumed to be 15\%.
} of an implanted blood glucose sensor that usually lasts for 254 days with RF links ($2\,\mathrm{mW}$~\cite{RF}) will extend upto 300 days. The simulation and empirical results in Fig.\ref{fig:exprplot} depicts the accurate matching of the simulation model tailored for the dimensions and properties of the porcine tissue with the empirical results. We next proceed with the simulation model for deeper analysis that includes a 3D tissue area of $100\times100\,\mathrm{cm^2}$ with the depth including skin, fat and muscle tissues embedding a maximum of $50$ iid nodes. 
 \begin{figure}[t]
 \centering
\includegraphics[width=8cm,height=3.6cm]
{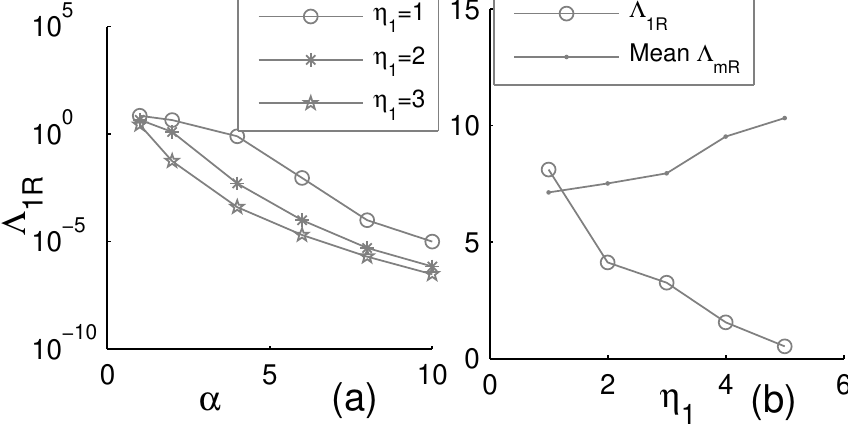}
\vspace{-2mm}
 \caption{(a) $\Lambda_{1R}$ vs $\alpha$ for varying $\eta$; (b) $\Lambda_{1R}$ \& mean $\Lambda_{mR}$ for varying $\eta$}
 \label{fig:4plota}
  \vspace{-3mm}
 \end{figure}

\noindent \textbf{S3. Optimized Link lengths:}  
In order to analyze the optimized inter-node distance obtained from fitted simulated results for both S-S \& M-S paths in the NICO phase, we compare it with that of the  expected $\Lambda_{mR_k}$ in ICAP phase (derived in Appendix) in Fig.\ref{fig:stat_results}. The ICAP distribution suggests longer $\Lambda_{mR_k}$ per cluster (mean $6.9\,\mathrm{cm}$) for $\Lambda^{th}=15\,\mathrm{cm}$. However, the fitted optimized distribution indicates shorter links with mean $\Lambda_{mR_k}\text{=}5.9\,\mathrm{cm}$ for S-S path with $\Lambda^{th}_{S-S}\text{=}15\,\mathrm{cm}$. The distribution of $\Lambda_{mR_k}$ for M-S path varies in accordance to its threshold. We note that for $\Lambda^{th}_{M-S}\text{=}20\,\mathrm{cm}$ that is higher than $\Lambda_{S-S}^{th}$, the mean M-S $\Lambda_{mR_k}$ is $4.1\,\mathrm{cm}$ that is significantly lower than the S-S path. Thus the algorithm efficiently minimizes $\Lambda_{mR_k}$ for M-S path even at higher thresholds. 

\begin{figure}[b]
 \centering
 \vspace{-4mm}
\includegraphics[width=9cm,height=3cm]
{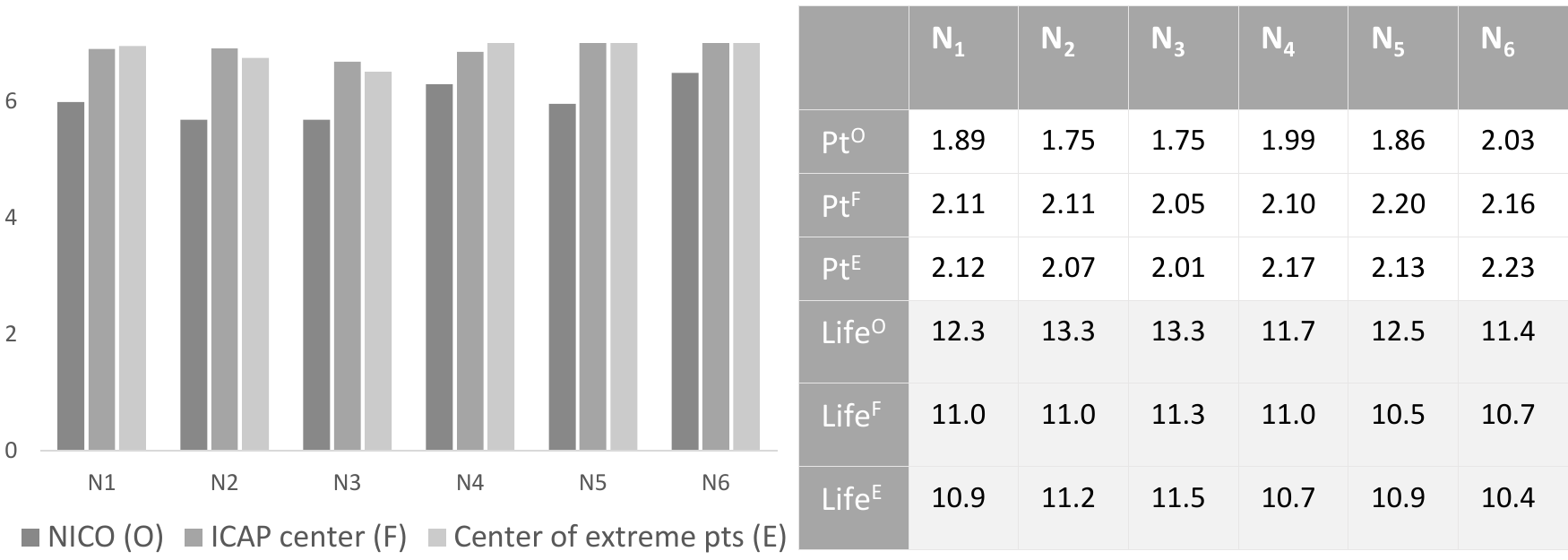}
\vspace{-4mm}
 \caption{Comparison of (left) link length, (right) $Pt(mW)$ and node life (in years) with relay position at $\hat{L}_{R_k}$, $L_{R_k}^F$ \& $L_{R_k}^E$ for S3}
 \label{fig:newplot}
 \end{figure}

\noindent \textbf{S4. All surface nodes cluster:} %We then analyze the link lengths in 6 node clusters with $I_k\text{=}0$ and uniform data rate. 
We compare our optimized relay position ($\hat{L}_{R_k}$) with conventional relay positions - at the ICAP cluster center ($L_{R_k}^{F}$) and at the center of extreme cluster node locations ($L_{R_k}^{E}$)~\cite{wsnbs} for 6 node clusters with $I_k\text{=}0$. Considering the $\sum \Lambda_{mR_k}$ over 50 simulations (refer Fig.\ref{fig:newplot}(left)), $\bar{\Lambda}_{mR_k}$ obtained with $\hat{L}_{R_k}$ is the lowest ($\bar{\Lambda}_{mR_k}\text{=}33.5\, \mathrm{cm}$, $\bar{\Lambda}_{mR_k}^{F}\text{=} 41.4\,\mathrm{cm}$ \& $\bar{\Lambda}_{mR_k}^E\text{=} 40.64\,\mathrm{cm}$). Thus, $\hat{L}_{R_k}$ gives $\approx 40\%$ more energy savings than at other positions. 

The average $Pt$ values (calculated for SNR of $5$ and $10\,\mathrm{KHz}$ bandwidth) of six S-S nodes are listed in Fig.\ref{fig:newplot}(right)). As expected, $\hat{L}_{R_k}$ offers extended life of upto $992$ days that is significantly higher than the conventional positions and hence becomes a critical component in GC-IBN topology design.

\noindent \textbf{S5. Single implant cluster:} We next consider a cluster of $6$ iid nodes of uniform data rates with $N_1$ implanted in muscle ($I_k\text{=}1$ at depth $z_1\text{=}0$)  and relay on surface at $\hat{L}_{R_k}$. The resulting $\Lambda_{1R_k}$ is significantly shorter than the other links as depicted in Table.\ref{tab:result2}. This ensures minimum $Pt_{1}$, as desired for an implant $N_1$ with a dramatic improvement of $88\%$ battery life  compared with the co-clustered nodes. %life extension of upto $75\,\mathrm{years}$. On average, an 
 \begin{figure}[t]
 \centering
\includegraphics[width=8cm,height=3.6cm]
{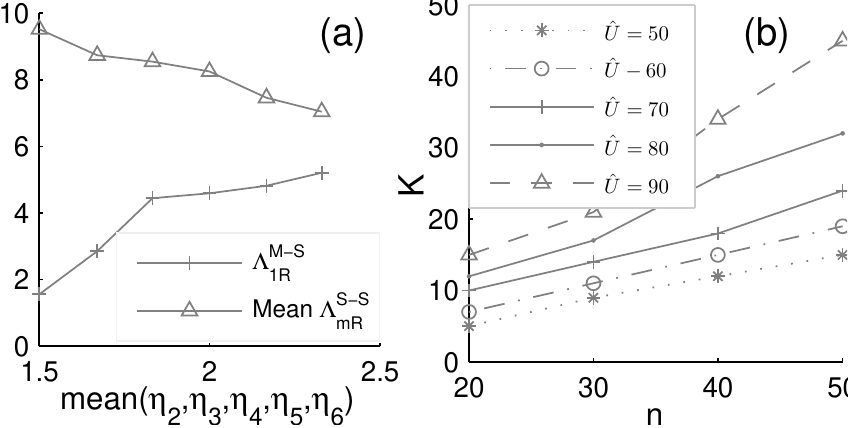}
\vspace{-2mm}
 \caption{(a) $\Lambda_{1R}$ vs mean $\eta_m$; (b) $n$ Vs $K$ for various $\hat{U}$}
 \label{fig:4plotb}
  \vspace{-3mm}
 \end{figure}
 
\noindent \textbf{Influence of $\alpha$:} Now we analyze the influence of $\alpha$ used in (\ref{eqn:wt}) on  $\Lambda_{mR_k}$. When $\alpha$ is reduced from $4$ to $2$, decreasing the priority given to conserving implant energy, the average $\Lambda_{1R_k}$ increases from $0.4\,\mathrm{cm}$ to $3.4\,\mathrm{cm}$ (refer Table.\ref{tab:result3}). When $\alpha$ is raised to $10$, the average $\Lambda_{1R_k}$ drops close to $0$. Thus, $Pt$ of implant can be controlled by varying $\alpha$. e.g., for the blood glucose sensor mentioned in S1, the extension of life can be upto 425 days when $\alpha$=$10$ that is $\approx 67\%$ more than a RF link and $\approx 40\%$ more than the relay position at $L_{R_k}^F$ with GC link. The influence of $\alpha$ is plotted with respect to $\eta$ in Fig.\ref{fig:4plota}(a).
\begin{table}[b]
\vspace{-5mm}
\centering
\caption{\label{tab:result2} Average $\Lambda_{mRk}$ (cm), $Pt_{m}$ (mW) \& node life (years) for S4}
\vspace{-3mm}
%\footnotesize
\scalebox{0.7}{
\begin{tabular}{ccccccc}
\toprule
&N1&N2&N3&N4&N5&N6\\
\midrule
$\Lambda_{mR_k}$ &0.4&	9.1&	9.0&	8.5&	7.8&	9.3\\
$Pt$&0.29&	2.73&	2.72&	2.59&	2.39&	2.79\\
Life&75.1&	8.5&	8.53&	8.96&	9.7&	8.31
\\
\bottomrule
\end{tabular}}
\vspace{-3mm}
\end{table}
\begin{table}[b]
\centering
\caption{\label{tab:result3} Influence of $\alpha$ over $\Lambda_{mR_k}$ for S4}
\vspace{-3mm}
\scalebox{0.7}{
\begin{tabular}{ccccccc}
\toprule
$\alpha$&N1&N2&N3&N4&N5&N6\\
\midrule
2&3.4&6.6&9.6&6.1&6.1&9.3\\
4&0.4&9.1&8.9&8.5&7.8&9.3\\
10&1E-3&13.3&15.6&14.4& 7.9&10.1\\
\bottomrule
\end{tabular}}
\end{table}	 

\noindent \textbf{Influence of $\eta$:} Considering the S5 scenario with $\alpha\text{=}4$, when $\eta_1$ increases from $1$ to $3$ (units), while $\eta_m\text{=}1,\,\forall m\text{=}\{2,..,6\}$, $\Lambda_{1R_k}$ is further reduced in accordance with (\ref{condn:eta}). Even when $N_1$ is on surface, when $\eta_1$ is increased from $1$ to $5$, the average $\Lambda_{1R_k}$ reduces from $8\,\mathrm{cm}$ to $0.3\,\mathrm{cm}$, penalizing the other links $\Lambda_{mR_k},m\text{=}\{2,..,6\}$ by $\approx 1.5\,\mathrm{cm}$ (Fig\,\ref{fig:4plota}(b)). Thus $Pt_{m}$ is reduced for higher $\eta$ towards energy balance. Similarly, for $\eta_1\text{=}5$, when the mean $\eta_m, m\text{=}\{2,..,6\}$, increases from $1$ to $2$, $\Lambda_{1R_k}$ is penalized from $1.6\,\mathrm{cm}$ to $4.8\,\mathrm{cm}$ (Fig.\ref{fig:4plotb}(a)), indicating a steeper decline with the rise in mean $\eta$, achieving equitable energy distribution.

\noindent \textbf{S6. Energy distribution in cluster with $I_k\textgreater 1$:} 
When there are multiple implants in a cluster, the energy consumption must be balanced. With $\hat{U}$ in (\ref{eqn:UB}) set to $90\%$ and the relay at $\hat{L}_{R_k}$, when the first implant is depleted of energy, the residual energy of other nodes in cluster varies between $4$ and $7\%$ ($\textless 10\%$ is desired). However, with the conventional relay positions at $L_{R_k}^F$ and $L_{R_k}^E$, the residual energy ranges between $14\%$ \& $18\%$, much higher than the desired level.
\begin{figure}[t]
 \centering
\includegraphics[width=5.8cm,height=4cm]{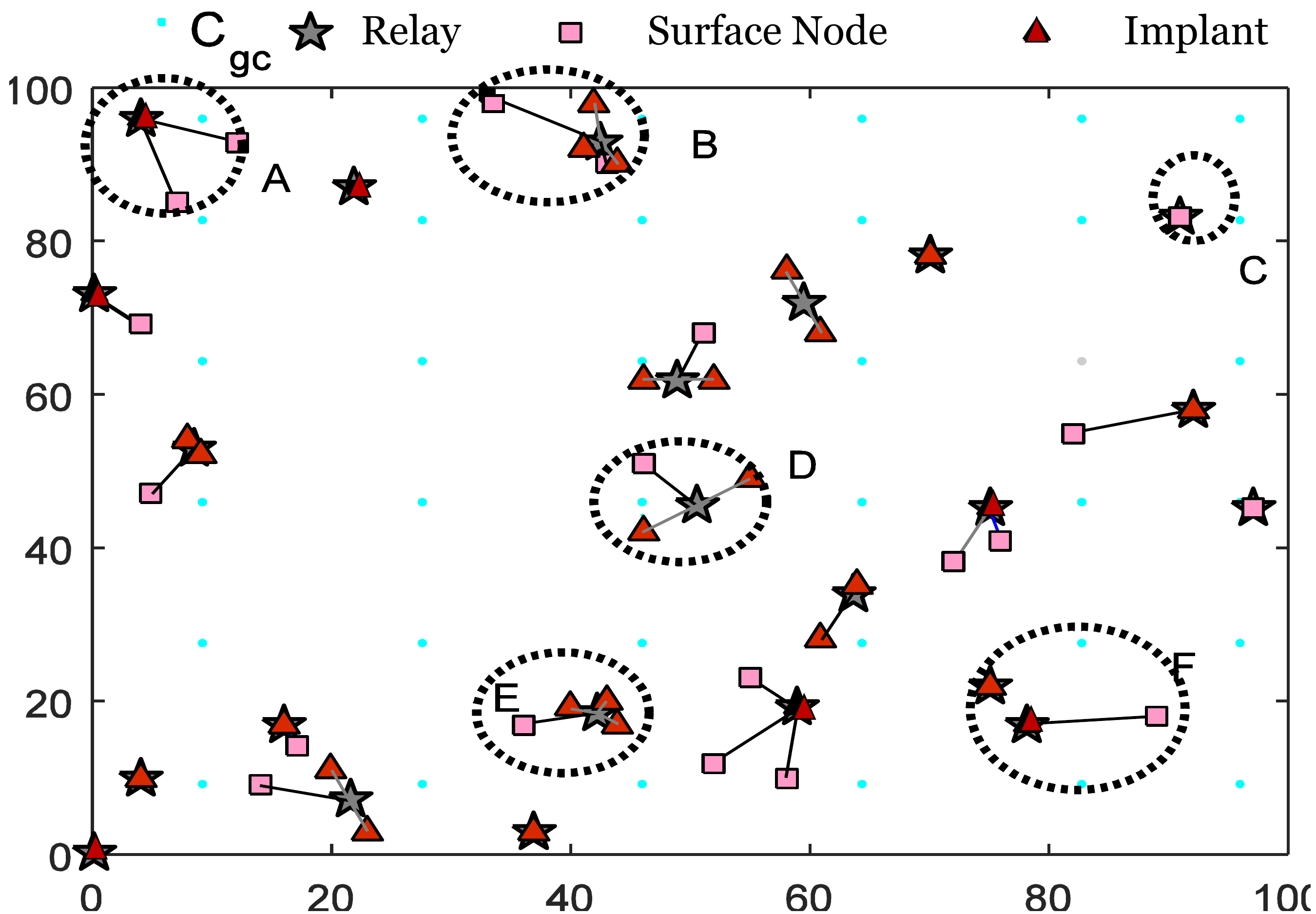}
 \vspace{-2mm}
 \caption{Optimized GC-IBN planar clusters}
 \label{fig:cluster2d}
  \vspace{-1mm}
 \end{figure}
 \begin{table}[t]
\centering
\caption{\label{tab:K} Average $K$ for $50$ iid nodes with $T_m=[1,2]$, $\eta_m\in [1,5]$, $\alpha\text{=}4$}
\vspace{-3mm}
\footnotesize
% \vspace{-4mm}
\begin{tabular}{ccccc}
\toprule
$\Lambda^{th},\Lambda^{th}_{S-S},\Lambda^{th}_{M-S}$& $8\,\mathrm{cm}$ & $10\,\mathrm{cm}$ & $12\,\mathrm{cm}$&$14\,\mathrm{cm}$\\
\midrule
$ICAP $ & 59 &51& 47& 45\\
$NICO $& 31& 25&22&18\\
\bottomrule
\vspace{-6mm}
\end{tabular}
\end{table}
\noindent \textbf{S7. Clustering efficiency:} %We evaluate our clustering efficiency in-terms of $K$ and link lengths.
Table.\ref{tab:K} illustrates the reduction in mean $K$ obtained from simulated NICO phase from that of ICAP (obtained as mean by fitting the $K$ distribution in Lemma\,\ref{sec:cluster}.1), for various threshold link lengths. %This reduction in cluster quantity is mainly due to (i) the ability to move the clusters to the position of the nodes, (ii) the ability to merge clusters and (iii) the flexibility of relay allocation only to those areas with nodes. 
$K$ intuitively increases with $n$ and also with $\hat{U}$ (Fig.\ref{fig:4plotb}(b)). For instance, for $\hat{U}\text{=}50\%$ and $n\text{=}50$, $K$ is only $15$ ($30\%$ of relays required). However, when $\hat{U}\text{=}90\%$ for the same $n$, $K$ becomes $45$, requiring $90\%$ of relays. Thus enforcing more uniformity in $Pt$ results in higher $K$.

The proposed framework satisfies the clustering objectives mentioned in Sec.\ref{sec:model} as shown below. In cluster of area $A$ (i.e., area under the dotted ellipse is $A$, as shown in Fig.~\ref{fig:cluster2d}), the relay is moved from $L_{R_A}^{F}$ to $\hat{L}_{R_A}$ towards the implant, reducing the link length by half. Clusters $D$ and $E$ demonstrate the shortening of the M-S links. Cluster $B$ reduces $K$ by merging multiple clusters. Cluster $C$ shows the assignment of dedicated relays for isolated nodes. The three nodes in the area enclosed by $F$ cannot be clustered together as the sum of their data rates ($\{4,4,5\}$) exceeds the cluster capacity ($Q_0\text{=}10$ units). Hence, another relay is assigned to the node with higher $\eta$ (5 units), based on the constraint in (\ref{eqn:clustersize}).	
\section{Conclusions}\label{sec:concl}
%While the invasive sensor technology and related medical applications advance rapidly, evolution in communication paradigm that safely transfers information through live tissue gains paramount importance. 
In this paper, we propose an energy efficient and energy balanced clustering framework suitable for galvanic coupled intra-tissue communication. The proposed framework comprises of two-phases of clustering that adapts to the signal propagation paths within heterogeneous tissues, and accommodates varying data rate requirements of implants. We demonstrate that the clustering approach not only minimizes the quantity and size of clusters, but also optimally positions the external signal pick up relays. Results indicate that our approach extends the lifetime of implants upto $70\%$ longer than the existing RF-based techniques. %\textcolor{blue}{XXX not sure what exists today...}also give some more concrete performance statistics... }. 

%In addition, the clusters with implants are handled specially towards energy balance. The intra-cluster topology framework proposed thus forms a critical part of GC-IBN design with key in-sights on tissue safety and reliability.

\section*{Acknowledgment} \label{ACK}
This material is based on the work supported by the U.S. 
National Science Foundation under Grant No. CNS-1453384. The authors are grateful for the suggestions from Rahman D. Mohammady \& Yousof Naderi, Northeastern University.
%\bibliographystyle{IEEEtran}
%\bibliography{IEEEabrv,bibs/p1,bibs/2,bibs/EB,bibs/article,bibs/newlit,bibs/clusterlinking}
\section*{Appendix: Proof of Theorem \ref{sec:cluster}.1}
\small
We assume that the coordinate distribution of nodes at a tissue $T$ follows two dimensional rectangular distribution over $[aF,\lambda+aF]$ horizontally and $[b\lambda,\lambda+b\lambda]$ vertically. Thus, the probability density functions of $X_m$, $X_{R_k}$, $Y_m$, $Y_{R_k}$ are given by 
\vspace{-2mm}
$f_{X_m}(x_m)\text{=}f_{X_{R_k}}(x_{R_k})\text{=}f_{Y_m}(y_m)\text{=}f_{Y_{R_k}}(y_{R_k}) = $
{\small
\begin{equation}
\begin{cases} \frac{1}{\lambda} &  x_m,  x_{R_K} \in [a\lambda,\lambda+a\lambda], y_m, y_{R_k} \in [b\lambda,\lambda+b\lambda] \\
0 & otherwise \end{cases}
\end{equation}}
%1/b-a
%while the cumulative distribution function is given by, 
%
%$F(x_m)\text{=}F(x_{R_k})\text{=}F(y_m)\text{=}F(y_{R_k})=$
%\begin{equation}
%\begin{cases} 0 &  x_m,  x_{R_K}, y_m, y_{R_k} \textless 0 \\
%	\frac{\{x_m,x_{R_K}, y_m, y_{R_k}\}}{\lambda} & \begin{cases}    a\lambda \textless x_m,x_{R_K} \textless  \lambda+a\lambda,\\ b\lambda \textless y_m, y_{R_k} \textless  \lambda+b\lambda \end{cases}\\
%1 & \begin{cases} x_m,  x_{R_K} \textgreater \lambda+a\lambda, \\ y_m, y_{R_k} \textgreater \lambda+b\lambda \end{cases} \end{cases}
%\end{equation}
%x-a/b-a
% mean of uniform distribution U[a,b] is 1/2(a+b) and variance is 1/12 (b-a)^2
%The expected value of X in (\ref{dist}), $E[X]$, is given by 
%\begin{equation} \label{eofx}
%\int_{a\lambda}^{\lambda+a\lambda} \int_{a\lambda}^{\lambda+a\lambda} |x_m\text{-}x_{R_k}| f_{X}(x_m,x_{R_k}) dx_m dx_{R_k}	
%\end{equation}
%where $f_{X_mX_{R_k}}(x_m,x_{R_k})$ is the joint probability density function of $X$, 
%that can be written as 
The joint pdf with $L_m$ and $L_{R_k}$ having independent coordinates is $
f_{X}(x)= f_{X_m}(x_m)f_{X_{R_k}}(x_{R_k})$. The corresponding CDF of X can be obtained by integrating $f_X(X)$ \cite{kostin}. $Y$ also has the similar distribution as that of $X$. Now the CDF of $Z=\sqrt{\Lambda}$ is obtained as $F_Z(z)= \iint_c f_X(x)f_Y(y)dx dy =$ 
{\small
\begin{equation}\label{eqn:jpd}
\begin{cases} 
0, &x\textless 0 \\
\pi \frac{x}{\lambda^2} -\frac{8}{3} (\frac{x}{\lambda^2})^{\frac{3}{2}} +\frac{1}{2}\frac{x^2}{\lambda^4}, & 0\leq x\textless \lambda^2\\
1-[\frac{2}{3}+2\frac{x}{\lambda^2}+\frac{1}{2}\frac{x^2}{\lambda^4}-\frac{2}{3}\sqrt{(\frac{x}
{\lambda^2}-1)^3} &\\
-2\sqrt{\frac{x}{\lambda^2}-1}(1-\frac{x}{\lambda^2}) -2\frac{x}{\lambda^2} sin^{-1}\frac{2-\frac{x}{\lambda^2}}{\frac{x}{\lambda^2}} ], & \lambda^2 \leq x \textless 2\lambda^2\\
1& X\geq 2\lambda^2
\end{cases}
\end{equation}
Assuming $\lambda'=\frac{r}{\lambda}$, the CDF of $\Lambda_{mR_k}$ in cluster $C_k$ becomes
\begin{equation}
F_{\Lambda}(r)=Pr(0\textless z \textless r^2)=\int_0^{r^2} f_Z(z) dz
\end{equation} 
$$=\begin{cases} 0, &r\textless 0\\
 \lambda'^2\pi-\frac{8}{3}\lambda'^3+\frac{1}{2}\lambda'^4,&0\leq r\textless \lambda\\
 1-[\frac{2}{3}+\frac{3}{2}\lambda'^2-\frac{2}{3}\sqrt{(\lambda'^2-1)^3}&\\
 -(2\sqrt{\lambda'^2-1})(1+\lambda'^2)-2\lambda'^2 asin\frac{2-\lambda'^2}{\lambda'^2}],&\lambda\leq r\textless \lambda\sqrt{2}\\
 1,&r\geq \lambda\sqrt{2}
 \end{cases}$$
%$F_{\Lambda_{max}}=[F_{\Lambda_{max}}(s)]^{N}$
%Assuming the left bottom corner in XY plane of cubical grid to be at origin and using (\ref{eqn:jpd}) in (\ref{eofx}), \begin{equation} \label{eofx2} % % %E[X]=\frac{1}{s^2} \int_{0}^{\lambda} \int_{0}^{\lambda} |x_m\text{-}x_{R_k}| dx_m .dx_{R_k} %\end{equation}
%%and similarly, 
%%\begin{equation} \label{eofy}
%%E[Y]=\frac{1}{s^2} \int_{0}^{\lambda} \int_{0}^{\lambda} |y_m\text{-}y_{R_k}| dy_m. dy_{R_k}
%\end{equation}
%Assuming $u=\sqrt{\Lambda}$, the pdf of the event ($u\leq\tau$) can be obtained as the convolution of $f_{X}$ in (\ref{eqn:jpd}) and $f_Y$ that is distributed similar to $f_X$ using (2.4.9) in \cite{geoprobability} as
%\begin{equation}
%f_u(\tau)=\begin{cases} 
%\int_0^{\tau} f_X(\tau\text{-}x)f_Y(x)dx, & 0\leq \tau \leq \lambda^2\\
%\int_{\tau-\lambda^2}^{\lambda^2} f_X(\tau\text{-}x)f_Y(x)dx, & \lambda^2\leq \tau \leq 2\lambda^2		   \end{cases}
%\end{equation}
%$$= \begin{cases} 
%\frac{\pi}{\lambda^2}-\frac{4\tau}{\lambda^3}+\frac{\tau}{\lambda^4}, & 0\leq \tau \leq \lambda^2\\
%\frac{4}{\lambda^2}[sin^{\text{-}1}(\frac{\lambda}{\sqrt{\tau}})\text{-}\frac{\pi}{4}\text{-}\frac{1}{2}]\text{+}\frac{4}{\lambda^3}\sqrt{\tau\text{-}\lambda^2}\text{-}\frac{\tau}{\lambda^4}, & \lambda^2\leq \tau \leq 2\lambda^2		   \end{cases}$$
%
%Now, the density of link length becomes, 
%\begin{equation}
%f_\Lambda(\Lambda)=f(u^2)\frac{du}{d\Lambda}=2\Lambda f_u(u^2)
%\end{equation}
The expected value of $\Lambda_{mR_k}$ in each cluster $E[\Lambda_{mR_k}]$ is 
%\begin{equation}\label{Eoflambda}
$$=\int_0^{\sqrt{2\lambda^2}} \Lambda_{mR_k} f_\Lambda(\Lambda_{mR_k}) d\Lambda_{mR_k}=\frac{\lambda}{3}ln(1+\sqrt{2})+\frac{\lambda\sqrt{2}}{15}(1+\sqrt{2})$$
%\end{equation}
\vspace{-4mm}
%\begin{center}
%$$=\frac{\lambda}{3}ln(1+\sqrt{2})+\frac{\lambda\sqrt{2}}{15}(1+\sqrt{2})$$
% \end{center} 
} 
 %$F_(x) = P(x_m-x_{R_k}\leq x) = $
%%with distribution (follows a triangular distribution)  
%\begin{equation}
%E[\bar{\Lambda}]= \int_0^\infty P(\bar{\Lambda} \textgreater \Lambda_{mR_k}) d\Lambda_{mR_k}
%\end{equation}

%\bibliographystyle{IEEEtran}
%\bibliography{IEEEabrv,bibs/p1,bibs/2,bibs/EB,bibs/article,bibs/newlit,bibs/clusterlinking}
\end{document}